\documentclass[aps,prl,twocolumn,amsfonts,showpacs,superscriptaddress,longbibliography]{revtex4-2}
\usepackage{verbatim,epsfig,amsmath,amssymb,bm,epsf,graphicx,psfrag,bbold,amsthm,amsfonts}
\usepackage[bottom]{footmisc}
\usepackage{hyperref}
\usepackage{cleveref} 
\usepackage{enumitem}
\usepackage{framed}
\usepackage{mathrsfs}
\usepackage{esint}
\setlist{nosep}
\usepackage{placeins}
\usepackage[all]{xy}
\usepackage{color}
\usepackage[utf8]{inputenc}
\usepackage{float}
\usepackage{natbib}
\usepackage{tikz-cd}
\usepackage{verbatim}
\usepackage{leftidx}
\usepackage[normalem]{ulem}

\usepackage{notes2bib}

\newcommand{\moy}[1]{\langle #1 \rangle}

\newcommand\beq {\begin{equation}}
\newcommand\eeq {\end{equation}}
\newcommand\beqa {\begin{equatiobn}\begin{array}}
\newcommand\eeqa {\end{array}\end{equation}}
\newcommand\bal {\begin{align}}
\newcommand\eal {\end{align}}
\newcommand{\bea}{\begin{eqnarray}}
\newcommand{\eea}{\end{eqnarray}}

\newcommand{\cD}{\mathcal{D}}

\theoremstyle{plain}

\theoremstyle{definition}

\theoremstyle{remark}

\begin{document}

\title{Universal spectral form factor for many-body localization}

\author{Abhishodh Prakash}
\email{abhishodh.prakash@icts.res.in (he/him/his)}
\affiliation{International Centre for Theoretical Sciences (ICTS-TIFR),
Tata Institute of Fundamental Research,
Shivakote, Hesaraghatta Hobli,
Bengaluru 560089, India}
\author{J. H. Pixley}
\email{jed.pixley@physics.rutgers.edu}
\affiliation{Department of Physics and Astronomy, Center for Materials Theory, Rutgers University, Piscataway, NJ 08854 USA}
\author{Manas Kulkarni}
\email{manas.kulkarni@icts.res.in}
\affiliation{International Centre for Theoretical Sciences (ICTS-TIFR),
Tata Institute of Fundamental Research,
Shivakote, Hesaraghatta Hobli,
Bengaluru 560089, India}

\date{\today}

\begin{abstract}
We theoretically study correlations present deep in the spectrum of many-body-localized  systems. An exact analytical expression for the  spectral form factor of Poisson spectra can be obtained and is shown to  agree well with numerical results on
two models exhibiting many-body-localization: a disordered quantum spin chain and a phenomenological $l$-bit model based on the existence of local integrals of motion. We also identify a  universal regime that is insensitive to the global density of states as well as spectral edge effects. 
\end{abstract}

\maketitle

\begin{figure}[t!]
	\centering
	\begin{tabular}{c}
		\includegraphics[width=0.47\textwidth]{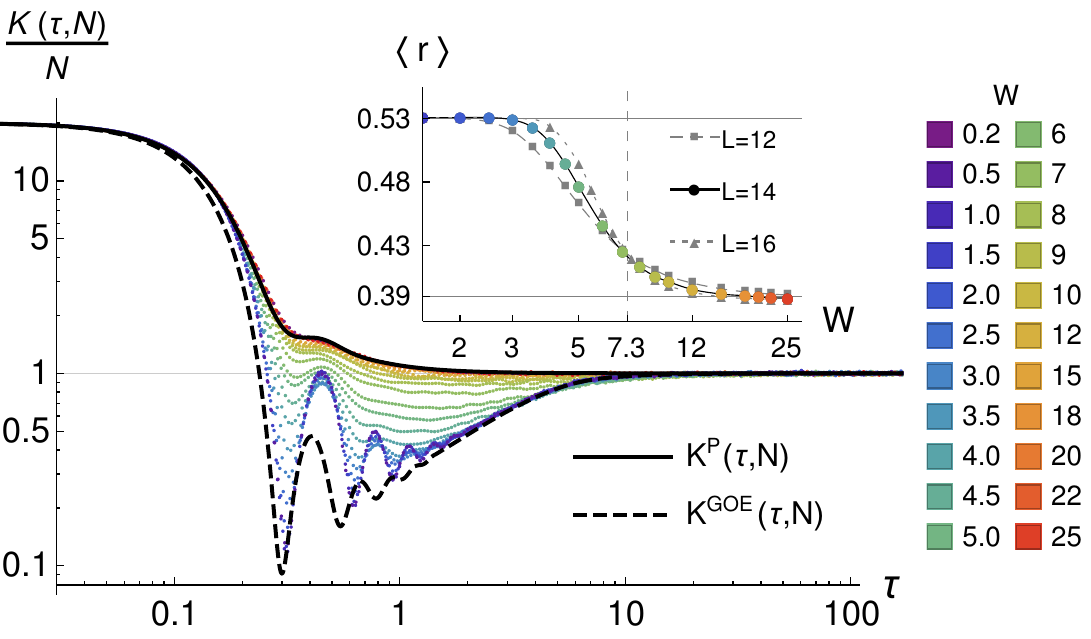}	
	\end{tabular}	
	\caption{\textbf{The spectral form factor  across the MBL transition.} This is  defined in \cref{eq:SFF} and computed from  $N=20$ eigenvalues  from the center of the many-body spectrum at different disorder strengths $W$ for the  Hamiltonian in~\cref{eq:Hamiltonian} with a system size $L=14$.  (Inset):~The adjacent gap ratio, $\langle r \rangle$ defined in \cref{eq:r} versus $W$. The approximate critical disorder, where the data at different system sizes cross is given by $W_c \approx 7.3$ has also been marked. For $W\gg W_c$ in the MBL phase the level statistics are Poisson $\langle r \rangle =2\ln(2) -1 \approx 0.39$ ~\cite{OganesyanHuse_2007_PhysRevB.75.155111}.  The dashed black line is the well known GOE expectation from random matrix theory known to describe the thermal phase, whereas the solid black line, that matches the numerical data in the MBL phase well (over the range of $W\ge 10$~\cite{supp}), is obtained in~\cref{eq:SFF MBL}. The analytical expressions $K^{GOE}(\tau,N)$ and $K^{P}(\tau,N)$ as well as the data are normalized to set the mean level spacing to unity.
	}
	\label{fig:r_Hamiltonian}
\end{figure}

\noindent Understanding how thermal equilibrium may, or may not emerge in isolated many-body quantum systems remains a central question in quantum statistical mechanics. \emph{Thermal} systems which are said to exhibit \emph{quantum chaos} satisfy the eigenstate thermalization hypothesis (ETH)~\cite{Srednicki_ETH_PhysRevE.50.888,Srednicki_1999}  whose subsystems equilibriate under their own dynamics.
In addition to being highly entangled, i.e. satisfying a ``volume law'' scaling with subsystem size, the eigenspectra of these systems exhibit long range repulsions that are captured by random matrix theory and produce universal features in their correlations measured in their \emph{spectral form factor} (SFF, defined below in \cref{eq:SFF,eq:SFFC}) such as the linear ramp~\cite{Haake_QuantumChaosBook,Cotler_SFFChaos2017,ShenkerGharibyan2018onsetofRM,Liu_SFFChaos_PhysRevD.98.086026,ChenLudwig_PhysRevB.98.064309,BertiniProsen_PhysRevLett.121.264101,KosLjubotinaProsen_PhysRevX.8.021062} (as shown in \cref{fig:r_Hamiltonian}). In the presence of strong quenched randomness or quasiperiodicity, quantum systems can become \emph{many body localized} (MBL)~\cite{HuseNandkishore2015_MBLreview,abanin2019colloquium,gopalakrishnan2020dynamics,deng2017many} where ETH is violated. In contrast to chaotic systems, MBL is characterized by eigenstates with short-range  ``area law'' entanglement and an absence of level repulsion. Recent experiments on ultra-cold atomic gases~\cite{schreiber2015observation,choi2016exploring,lukin2019probing}, trapped ions~\cite{Smith_2016_MBLExpt}, superconducting qubits~\cite{roushan2017spectroscopic,xu2018emulating} and nuclear spins~\cite{wei2018exploring} have provided evidence for the existence of the MBL phase.

Instabilities to MBL have been argued to arise in high dimensions~\cite{DeRoekHuveneers_bubble_PhysRevB.95.155129} and in the presence of certain symmetries~\cite{PotterVasseur_PhysRevB.94.224206}. More recently however, the very existence of the MBL phase has been challenged based on a finite size scaling analysis of the linear ramp  of the SFF on approach to the MBL transition from the chaotic side~\cite{Prosen_2019arXiv190506345S}. A critique of this work was subsequently presented~\cite{Abaninetal_ReplyToProsen_2019arXiv191104501A} pointing out the intricacies involved in finite sized calculations and conclusions drawn from them, while further studies have highlighted the difficulty in studying the MBL transition in finite size numerics~\cite{sierant2020thouless, panda2020can}. Recently, the authors of Ref.~\cite{Prosen_2019arXiv190506345S} pointed out that their claim of the absence of MBL is due to their choice of scaling function, which instead should follow a Kosteritz-Thouless ``like'' scaling form as they demonstrate in Ref.~\cite{vsuntajs2020ergodicity} consistent with recent theories of the MBL transition~\cite{goremykina2019analytically,dumitrescu2019kosterlitz,morningstar2020many}. 
Irrespective of the question of validity of the finite-size numerics in the vicinity of the MBL \emph{transition}, the question of how to characterize the  MBL \emph{phase} using the SFF alone, is undoubtedly worthy of further examination. 
If the MBL phase indeed exists, it is conceivable that its SFF has its own universal features to which any putative system exhibiting MBL should be compared. However, apart from a few hints~\cite{ChanLucaChalker_PhysRevLett.121.060601}, the existence of such a form and an understanding of its features has been lacking thus far.

In this Letter, we  investigate the spectral correlations in MBL systems. We show that the SFF for MBL systems can be calculated deep in the spectrum due to their convergence with \emph{Poisson levels} for which we can derive an exact analytical expression with a finite number of levels~\cite{riser2020nonperturbative}, (plotted as a solid line in~\cref{fig:r_Hamiltonian}).  
We determine the validity of this expression by comparing it with numerical simulations of a phenomenological $l$-bit model~\cite{SerbinPapicAbanin_lbit_PhysRevLett.111.127201,HuseOganesyan_lbit_PhysRevB.90.174202} as well as a microscopic disordered many-body Hamiltonian. In both cases, by focusing on states in the middle of the many body spectra where the many-body density of states is nearly flat, we find excellent agreement between the exact expression and the numerical results. In the limit of an infinite number of levels, to leading order, our results reduce to the general expectation of integrable systems due to Berry and Tabor~\cite{BerryTabor1977_level,BerryTabor_conjecture_Jens}. However, we show that the leading  correction to the SFF beyond that of Berry and Tabor is universal in an intermediate power-law scaling regime and is robust to changes in the global density of states as well as spectral edge effects. Our results provide further support for the existence of the MBL phase in one dimension.

\medskip

\emph{Models for many-body localization}: To make a detailed comparison with the properties of the MBL phase, we consider two different models. The first is a quantum spin-chain with quenched disorder whose Hamiltonian is
\begin{multline}
\label{eq:Hamiltonian}
H = \sum_i   J_1 (S^x_i S^x_{i+1} + S^y_i S^y_{i+1} + \Delta S^z_i S^z_{i+1})+  w_i S^z_i \\ +  \sum_i J_2 (S^x_i S^x_{i+2} + S^y_i S^y_{i+2} + \Delta S^z_i S^z_{i+2}). 
\end{multline}
$S^\alpha$ are spin operators that can be written in terms of Pauli matrices as $S^\alpha= \frac{1}{2} \sigma^\alpha$ and the random couplings $w_i$ are drawn from a uniform distribution $[-W, W]$.
Variants of this model have been previously studied~\cite{Prosen_2019arXiv190506345S,KhemaniShengHuse_CriricalMBL_PhysRevLett.119.075702,khemani2017critical} and are known to have a thermal phase at weak disorder and an MBL phase at  strong disorder.
Following Ref~\cite{Prosen_2019arXiv190506345S}, we set $J_1 = J_2 = 1.0$ and $\Delta = 0.55$.  

Deep in the MBL phase, any local Hamiltonian such as \cref{eq:Hamiltonian} can be described by a complete set of emergent local integrals of motion~\cite{SerbinPapicAbanin_lbit_PhysRevLett.111.127201,HuseOganesyan_lbit_PhysRevB.90.174202}. This means that there should exist a finite depth unitary circuit $U$ that can recast $H$ into a diagonal form, $U H U^\dagger = H_{lbit}$: 
\begin{equation}
\label{eq:lbit Hamiltonian}
H_{lbit} = \sum_i J^{(1)}_{i} \kappa^z_i + \sum_{i,j} J^{(2)}_{ij} \kappa^z_i \kappa^z_j + \sum_{i,j,k} J^{(3)}_{ijk} \kappa^z_i \kappa^z_j \kappa^z_k + \ldots
\end{equation}
where $\kappa^z_i$ are the so called $l$-bit Pauli operators with localized support on the Hilbert space near site $i$, whose eigenvalues represent the locally conserved quantities and the magnitudes of $J^m_{i_1 \ldots i_m}$ fall off exponentially with distance.
The second model we consider is a truncated version of the above phenomenological $l$-bit model~\cref{eq:lbit Hamiltonian}  where we retain only up to 10 spin nearest neighbor interactions with all couplings drawn from a uniform distribution $J^{(1 \ldots 10)} \in [-1,1]$. 

\FloatBarrier

\medskip

\emph{Characterizing spectral correlations of quantum systems}: A popular diagnostic used to distinguish MBL and chaotic systems via their spectral correlations is the \emph{adjacent gap ratio} ($r$) ~\cite{OganesyanHuse_2007_PhysRevB.75.155111}. This is defined in terms of successive gaps $\delta_i = E_{i+1} - E_i$ of an ordered energy spectrum $\{E_i\}$ as follows:
\begin{equation}
r_i = \frac{\mathrm{min}(\delta_i, \delta_{i+1})}{\mathrm{max}(\delta_i, \delta_{i+1})}.
\label{eq:r}
\end{equation}
For chaotic systems, the value of $\moy{r}$ (where $\moy{\dots}$ denotes averaging over samples and energy) can be computed from an appropriate random matrix ensemble. For example, the Gaussian orthogonal ensemble (GOE), appropriate for systems with time-reversal symmetry  gives $\moy{r} \approx 0.53$, while Poisson levels, applicable for MBL systems gives $\moy{r} =2\ln(2)-1 \approx 0.39$~\cite{OganesyanHuse_2007_PhysRevB.75.155111}. As shown in the inset of~\cref{fig:r_Hamiltonian}, by tracking $\moy{r}$, we can see that the Hamiltonian of \cref{eq:Hamiltonian} supports a thermal phase for small $W$ and an MBL phase for large $W$ with the critical disorder strength somewhere near $W_c \approx 7.3$ where the curves for different system sizes cross, consistent with previous work~\cite{Prosen_2019arXiv190506345S}. 

The adjacent gap ratio captures the repulsion of neighboring levels, and thus only probes \emph{local} spectral correlations. It does not measure long-range spectral correlations, which have important and useful information. 
A more 
comprehensive
diagnostic 
is the spectral form factor (SFF)~\cite{Haake_QuantumChaosBook}, which is the main tool of analysis in this Letter and is defined in terms of eigenvalues $\{E_i\}$ as follows:
\begin{equation}
K(\tau,N) =  \langle \sum_{m,n=1}^N e^{i \tau (E_m - E_n)} \rangle 
\label{eq:SFF}
\end{equation}
where $N$ is the total number of eigenvalues in consideration. Also useful is the connected SFF defined as
\begin{equation}
K_c(\tau,N) =  \langle \sum_{m,n=1}^N e^{i \tau (E_m - E_n)}   \rangle - |\langle \sum_{m=1}^N e^{i \tau E_m}   \rangle|^2.
\label{eq:SFFC}
\end{equation}
The information about long-range correlations is contained in the form of $K(\tau,N)$ interpolating the early and late $\tau$ values of $N^2$ and $N$ respectively ($0$ and $N$ for $K_c(\tau,N)$). For chaotic systems, just like $\moy{r}$, the SFF can also be computed from an appropriate random matrix ensemble. For instance, as seen in \cref{fig:r_Hamiltonian}, the SFF for the Hamiltonian~\cref{eq:Hamiltonian} with weak disorder strength ($W$)   exhibits a clear ramp and matches that of the GOE ensemble whose approximate expression (plotted as a dotted line) is  known~\cite{Haake_QuantumChaosBook,Cotler_SFFChaos2017,ShenkerGharibyan2018onsetofRM,Liu_SFFChaos_PhysRevD.98.086026,ChenLudwig_PhysRevB.98.064309,BertiniProsen_PhysRevLett.121.264101,KosLjubotinaProsen_PhysRevX.8.021062}

As we increase the disorder strength, as shown in~\cref{fig:r_Hamiltonian}, the SFF qualitatively changes as the model passes through the MBL transition with the disappearance of the ramp. 
Deep in the MBL phase (i.e. where $\moy{r}\approx0.39$), the SFF again takes on a new stable form $K^{P}(\tau,N)$ (plotted as a solid black line in fig.~\ref{fig:r_Hamiltonian}). The expression for $K^{P}(\tau,N)$ as well as the connected version, $K_c^{P}(\tau,N)$ we obtain are presented in \cref{eq:SFF MBL,eq:SFFc MBL}. We will show in the following section that they correspond to energy levels drawn from a \emph{Poisson process}. 

Contrasting features between $K^{GOE}$ and $K^{P}$ can be seen at intermediate $\tau$ values, in the regime where the SFF is expected to be universal (this occurs in the range $\frac{1}{\mu \cD} \lesssim \tau \lesssim 1/\mu$ for $K^{GOE}$ and $\frac{1}{\sqrt{\mu \cD}} \lesssim \tau \lesssim 1/\mu$ for $K^{P}$~\cite{supp}) where $\mathcal{D} = \mu N$ is the  many-body bandwidth of the chosen levels and $\mu$ is the  mean level spacing. For $K^{GOE}$, this corresponds to the `ramp' region~\cite{Haake_QuantumChaosBook,Cotler_SFFChaos2017,ShenkerGharibyan2018onsetofRM,Liu_SFFChaos_PhysRevD.98.086026,ChenLudwig_PhysRevB.98.064309,BertiniProsen_PhysRevLett.121.264101,KosLjubotinaProsen_PhysRevX.8.021062}). On the other hand, as expected, $K^{P}(\tau,N)$ lacks the ramp but exhibits a universal sub leading power-law form that will be discussed later.  

\medskip
\emph{SFF for Poisson levels}:
\label{sec:Two possibilites for MBL SFF}
The single-particle spectrum of the Anderson insulator~\cite{AndersonLocalization_PhysRev.109.1492} deep in the localized phase can simply be described by a set of uncorrelated random numbers (the values of random chemical potentials). In this case, on scales smaller than the single-particle bandwidth, the spectrum looks like a \emph{Poisson process}~\cite{mehta2004random,Haake_QuantumChaosBook}. For example, the distribution of the level spacings is exponential $P(\delta) = \frac{1}{\mu} \exp \left(-\frac{\delta}{\mu}\right)$. Consistent with the hypothesis of emergent integrability in localized systems, this is identical to the distribution of level spacings in point particle systems with integrable classical trajectories conjectured by Berry and Tabor~\cite{BerryTabor1977_level,BerryTabor_conjecture_Jens} and has been verified in several systems~\cite{Casati_billiards_PhysRevLett.54.1350,marklof1998spectral_billiards}.

\begin{figure}[!htbp]
	\centering
	\begin{tabular}{c}
		\includegraphics[width=0.45\textwidth]{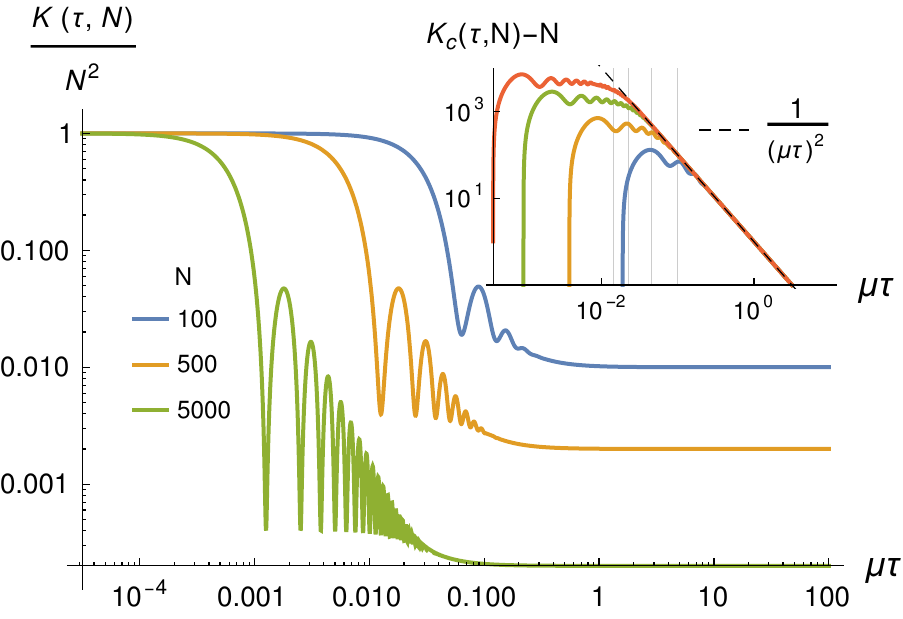}	\\
		\includegraphics[width=0.45\textwidth]{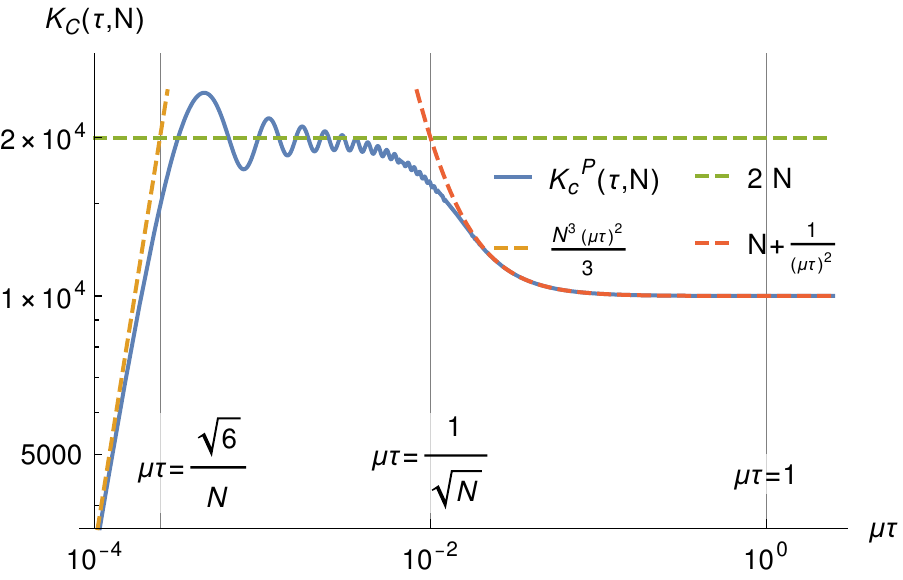}			
	\end{tabular}	
	\caption{{\bf SFF for Poisson levels} (Above:) The SFF for Poisson levels for various values of $N$. The reduced SFF $K_c(\tau) - N$ exposes the universal form (dashed lines) which sets in after a time $ \mu  \tau = \frac{1}{ \sqrt{N}}$ (marked for each $N$) is shown in the inset. (Below:) The various universal and non-universal $\tau$ regimes are shown for the connected SFF. 
	\label{fig:SFF_analytical}}
\end{figure}

The many body levels of the Anderson insulator on the other hand is a weighted sum of the single particle eigenvalues. For a system of size $L$, the $\sim \mathcal{O}(L)$ random numbers present in the Hamiltonian are used to generate $\sim 2^L$ many-body eigenvalues and are no longer completely uncorrelated. How the spectrum further changes in the presence of interactions for MBL systems is less obvious. However, extensive work~\cite{SerbynMoore_PhysRevB.93.041424,OganesyanHuse_2007_PhysRevB.75.155111,LevelstatistticsMBL_PhysRevB.99.104205,RMT_MBL_PhysRevLett.122.180601,RMT_MBL_PhysRevB.101.104201} has provided evidence that the Poisson nature continues to persist in the many-body levels of MBL systems on energy scales smaller than the many-body bandwidth~\cite{OganesyanHuse_2007_PhysRevB.75.155111}. To compute the SFF, we need more information than local statistics such as the level spacing distribution - we need the joint distribution of eigenvalues $P(E_n,n;E_m,m)$  i.e. the likelihood of the $n^{th}$ level to be $E_n$ when the $m^{th}$ level is $E_m$. For Poisson process, this is~\footnote{It should be noted that using the joint distribution for uncorrelated levels i.e. $P(E_n,n;E_m,m) = P(E_m) P(E_n)$ would produce a different SFF. See the supplementary materials~\cite{supp} for more details about this and the relationship to the SFF form in the main text.}~\cite{supp} (assuming $m>n$)
\begin{align}
P(E_n,n;E_m,m) &= p(E_n,n)~ p(E_m-E_n,m-n)
\end{align}
where $p(E_k,k)$ is the well known Poisson distribution
\begin{equation}
p(E_k,k) =   \frac{e^{-\frac{E_k}{\mu} }}{\mu (k-1)!} \left(\frac{E_k}{\mu}\right)^{k-1} . 
\end{equation}
Using this, we can exactly obtain the expressions for the Poisson SFF~\cite{supp}. 
\begin{align}
K^{P}(\tau,N) &= N + \frac{2}{(\mu \tau)^2}  - \frac{ (1+i \mu \tau)^{1-N} + (1-i \mu \tau)^{1-N}  }{(\mu\tau)^2} \label{eq:SFF MBL} \\
K^{P}_c(\tau,N) 
&= N + \frac{1}{(\mu \tau)^2} - \frac{(1+(\mu \tau)^2)^{-N}}{(\mu \tau)^2} \nonumber  \\ & ~~~~~~~~~~~ - \frac{i}{  \mu \tau } \left[ (1 + i  \mu \tau )^{-N} - (1 - i \mu  \tau )^{-N}  \right] \label{eq:SFFc MBL}
\end{align}
Note that these expressions have been also obtained by the authors of \cite{riser2020nonperturbative} as a special case of a more general result applicable to spectra with uncorrelated gaps. Our focus is on the application of these results to the MBL spectrum where the Poisson nature is emergent and not intrinsic.We now proceed to understand various limiting regimes of the above expressions. In the limit of $N\rightarrow \infty$ we obtain the expected result of Berry-Tabor $\lim_{N\rightarrow \infty} K^P(\tau,N)/N=1+\delta(\tau)$. 
If $\cD = \mu N$ is the bandwidth of the selected eigenvalues with mean level spacing $\mu$, the early $\tau$ behavior ($ \tau < \frac{1}{\sqrt{\mu \cD}}$) is largely determined by the Poisson density of states (DOS)~\cite{supp} which is non-trivial only at the edges, and is not a universal feature. Just as in the case of RMT, the interesting part is at intermediate values of $\tau$ i.e. $\frac{1}{ \sqrt{ \mu \cD}} < \tau < \frac{1}{\mu}$ (see \cref{fig:SFF_analytical} and \cite{supp} for details of various universal and non-universal $\tau$ regimes), where we have 
\begin{equation}
K^P_c(\tau,N) = N + \frac{1}{(\mu \tau)^2}+\mathcal{O} \left(\frac{1}{N}\right),
\label{eq:SFF reduced}
\end{equation}
 and the disconnected part behaves similarly $K^P(\tau,N)-N\sim 2/(\mu\tau)^2$.
The leading $N$ is merely the large $\tau$ value and is frequently quoted as the SFF signature of Poisson spectra. More interesting is the sub leading $\frac{1}{(\mu \tau)^2}$ term that is  $N$ independent.
This suggests that if we subtract the dominant trivial value and consider $K_c(\tau,N)-N$ which we dub the \emph{reduced SFF}, it should assume a $\frac{1}{(\mu \tau)^2}$ form that survives the $N \rightarrow \infty$ limit and is \emph{universal} in the same way that the ramp is universal to RMT i.e. the form is robust to effects from spectral edges arising from a finite bandwidth as well as non-trivial global density of states~\cite{Cotler_SFFChaos2017}. Note that similar time-scales as well as scaling forms were also obtained in \cite{riser2020nonperturbative} even though their notion of universality (independence of underlying gap distribution) is  different from ours~\cite{supp}. We verify this using the physical models mentioned before where non-negligible edge and DOS effects are expected.
 
\emph{Comparison with numerical calculations}:
We now numerically check the analytical results of the previous section by focusing on the two models defined in~\cref{eq:Hamiltonian} and \cref{eq:lbit Hamiltonian}.
Both models possess a global $U(1)$ spin rotation symmetry which allows us to focus on  half-filling i.e. the total $S^z = 0$ sector. We will perform our analysis by shifting the $N$ chosen eigenvalues by the smallest one so as to make them non-negative. For ease of comparison with the analytical results as well as across system sizes, after averaging over disorder samples, we re-scale $\tau$ by the mean level spacing $\mu$ effectively setting $\mu=1$. Depending on system size, our analysis is performed using disorder samples ranging from $10,000$ to $50,000$~\cite{supp}.

\begin{figure}[t!]
	\centering
	\begin{tabular}{c}
		\includegraphics[width=0.45\textwidth]{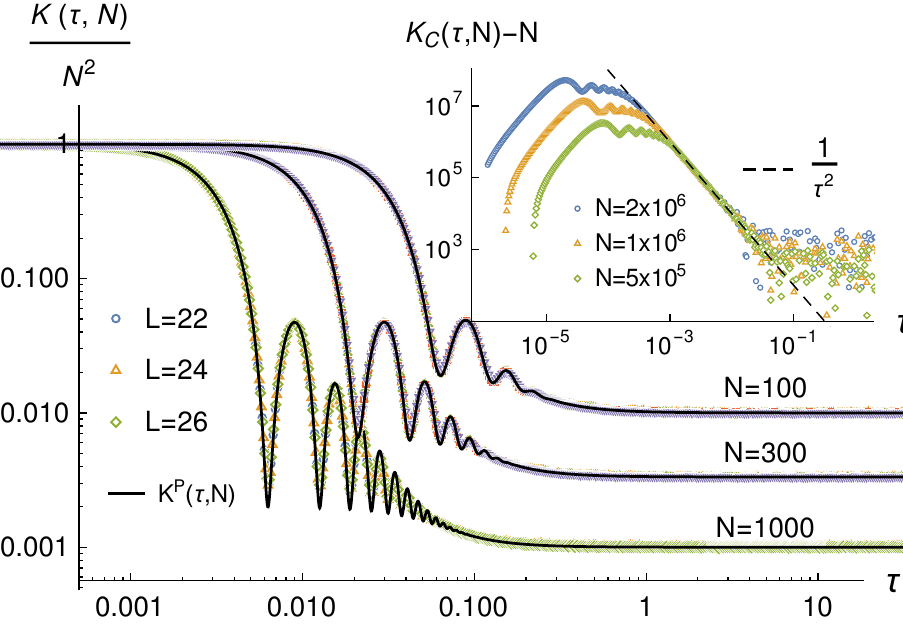} 	 \\
		\includegraphics[width=0.45\textwidth]{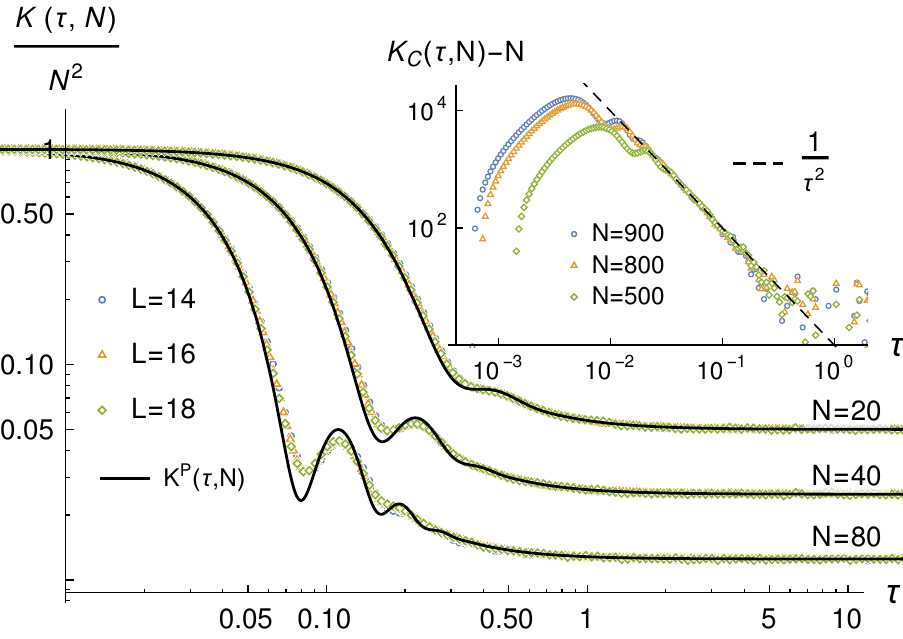} 		
	\end{tabular}
	\caption{{\bf Comparing the SFF for Poisson levels with models of MBL}. The SFF for the $l$-bit model of~\cref{eq:lbit Hamiltonian} (above) and the microscopic Hamiltonian in~\cref{eq:Hamiltonian}  deep in the MBL phase with $W=25$  (below)  for various system sizes ($L$) and small numbers of eigenvalues ($N$) drawn from the middle of the many-body spectrum compared with the analytical curves $K^P(\tau,N)$ of~\cref{eq:SFF MBL}. Deviations appear at short $\tau$ but are absent in the universal regime at intermediate $\tau$.  Reduced SFF shown for large values of $N$ for the $L=26$ $l$-bit model (inset, above) and $L=18$ Hamiltonian $H(W)$ (inset, below) that clearly demonstrates the universal $1/\tau^2$ behavior at intermediate $\tau$. 
	}
	\label{fig:SFF_Lbit}
\end{figure}

It is a well known challenge to compare exact random matrix theory predictions with numerics on microscopic models due to the difference in their DOS, particularly at the edges of the spectral bandwidth. The early $\tau$ behavior of the SFF in particular deviates from the RMT prediction due to this and a better agreement can be obtained by a careful \emph{unfolding} of the spectrum~\cite{JiaVerbaaschot_2020,Haake_QuantumChaosBook}. However, as the authors of Ref~\cite{ShenkerGharibyan2018onsetofRM} point out, at intermediate values of $\tau$, the ramp is robust to these effects and can be observed even without unfolding. Coming to our Poisson case, the situation is similar - the early $\tau$ behavior is affected by the overall DOS of the microscopic models and thus deviates from the analytical form of $K^P(\tau,N)$  in \cref{eq:SFF MBL}. For a fixed number of eigenvalues $N$, these deviations are reduced by increasing the system size $L$ (and thus the total Hilbert space $\mathcal{N}_L$). In the thermodynamic limit ($L \rightarrow \infty$) when the parameter
$\zeta \equiv N/\mathcal{N}_L$  vanishes for any finite $N$, we expect any deviations to completely vanish and the analytical results to match exactly~\cite{supp}. Nevertheless, as suggested previously, even for large values of $N$ when the early $\tau$ form deviates significantly, the SFF matches at intermediate-$\tau$ values where the SFF is universal and is best seen by in the reduced SFF, $K_c(\tau,N)-N$.

We start with the $l$-bit model of~\cref{eq:lbit Hamiltonian}. Since it is already diagonal, the eigenvalues are generated easily and as a result, we are able to reach relatively large system sizes. As seen in ~\cref{fig:SFF_Lbit} (top panel), the numerical SFF, $K(\tau,N)$ matches the analytical one for Poisson levels, $K^P(\tau,N)$ of \cref{eq:SFF MBL} (dotted lines) very well with negligible deviations for small values of $N$. For $N \sim 1000$, deviations start becoming visible at short-$\tau$.  The universal intermediate-$\tau$ form is very clearly seen at large $N$  in the reduced SFF (inset, top panel) as this expands the universal temporal regime $\frac{1}{ \sqrt{ \mu \cD}} < \tau < \frac{1}{\mu}$.  

We now turn to the microscopic Hamiltonian~\cref{eq:Hamiltonian} and focus deep in the MBL phase at $W=25$, where $\moy{r}\approx 0.39$ is nicely Poisson at the accessible $L$. Here, we are relatively limited in the system sizes that we can reach and the presence of complex microscopic details further impacts the finite sized numerical results more severely than in the case of the idealized $l$-bit model. Nevertheless, as seen in~\cref{fig:SFF_Lbit} (lower panel), for small values of $N$ (20,40), the numerical SFF matches the analytical equation \cref{eq:SFF MBL} (dotted lines) very well. For larger values of $N \sim 80$, deviations start becoming visible at short $\tau$ values. Again, the universal intermediate-$\tau$ form is very clearly seen at large $N$ (inset, bottom panel). Although we have only presented the analysis for $W=25$, we find that all these results remain virtually unchanged for a wide range of disorder strengths, $W\ge 10$~\cite{supp}. This strongly supports the notion that MBL is a robust phase in disordered one-dimensional isolated quantum many-body systems.

\medskip 
\emph{Conclusion}:
In this Letter, we have derived an exact expression for the spectral form factor of Poisson levels and identified a universal regime. We have shown that this describes the SFF in the many body localized phase well through a detailed comparison with numerical results on two separate physical models. The analytic expression of the spectral form factor obtained here is expected to apply to any integrable many-body quantum system. In particular, we conjecture that in the SFF of integrable models, the universal power-law correction should be observed as a \emph{refined} version of the Berry-Tabor conjecture.

\medskip
\emph{Note added:} During the later stages of our work, we became aware of a recent mathematical physics paper~\cite{riser2020nonperturbative} which also comprehensively discusses the spectral form factor for spectra with uncorrelated spacings in a distinct context. At the time of submission, we were also made aware of ref~\cite{riser2020power} which has some overlapping results presented in the Supplemental Materials~\cite{supp}. 

\medskip 
\emph{Acknowledgments}: We acknowledge helpful discussions with Anirban Basak, Po-Yao Chang, Arghya Das, Subroto Mukerjee, Tomaz Prosen,  Sthitadhi Roy, and Marko Znidaric. This work was initiated  at the Aspen Center for Physics, which is supported by National Science Foundation Grant No. PHY-1607611.
A.P. receives funding from the Simon's foundation through the ICTS-Simons prize postdoctoral fellowship. M.K. acknowledges support via a Ramanujan Fellowship
(SB/S2/RJN-114/2016), an Early Career Research Award (ECR/2018/002085) and a Matrics Grant (MTR/2019/001101) from the Science and Engineering Research Board (SERB), Department of Science and Technology, Government of India. A.P. and M.K. acknowledge support of the Department of Atomic Energy, Government of India, under Project No. RTI4001. J.H.P. is partially supported by  NSF Career Grant No. DMR-1941569.

 \FloatBarrier

\bibliography{references}{}

\begin{thebibliography}{51}%
\makeatletter
\providecommand \@ifxundefined [1]{%
 \@ifx{#1\undefined}
}%
\providecommand \@ifnum [1]{%
 \ifnum #1\expandafter \@firstoftwo
 \else \expandafter \@secondoftwo
 \fi
}%
\providecommand \@ifx [1]{%
 \ifx #1\expandafter \@firstoftwo
 \else \expandafter \@secondoftwo
 \fi
}%
\providecommand \natexlab [1]{#1}%
\providecommand \enquote  [1]{``#1''}%
\providecommand \bibnamefont  [1]{#1}%
\providecommand \bibfnamefont [1]{#1}%
\providecommand \citenamefont [1]{#1}%
\providecommand \href@noop [0]{\@secondoftwo}%
\providecommand \href [0]{\begingroup \@sanitize@url \@href}%
\providecommand \@href[1]{\@@startlink{#1}\@@href}%
\providecommand \@@href[1]{\endgroup#1\@@endlink}%
\providecommand \@sanitize@url [0]{\catcode `\\12\catcode `\$12\catcode
  `\&12\catcode `\#12\catcode `\^12\catcode `\_12\catcode `\%12\relax}%
\providecommand \@@startlink[1]{}%
\providecommand \@@endlink[0]{}%
\providecommand \url  [0]{\begingroup\@sanitize@url \@url }%
\providecommand \@url [1]{\endgroup\@href {#1}{\urlprefix }}%
\providecommand \urlprefix  [0]{URL }%
\providecommand \Eprint [0]{\href }%
\providecommand \doibase [0]{https://doi.org/}%
\providecommand \selectlanguage [0]{\@gobble}%
\providecommand \bibinfo  [0]{\@secondoftwo}%
\providecommand \bibfield  [0]{\@secondoftwo}%
\providecommand \translation [1]{[#1]}%
\providecommand \BibitemOpen [0]{}%
\providecommand \bibitemStop [0]{}%
\providecommand \bibitemNoStop [0]{.\EOS\space}%
\providecommand \EOS [0]{\spacefactor3000\relax}%
\providecommand \BibitemShut  [1]{\csname bibitem#1\endcsname}%
\let\auto@bib@innerbib\@empty
\bibitem [{\citenamefont {Srednicki}(1994)}]{Srednicki_ETH_PhysRevE.50.888}%
  \BibitemOpen
  \bibfield  {author} {\bibinfo {author} {\bibfnamefont {M.}~\bibnamefont
  {Srednicki}},\ }\bibfield  {title} {\bibinfo {title} {Chaos and quantum
  thermalization},\ }\href {https://doi.org/10.1103/PhysRevE.50.888} {\bibfield
   {journal} {\bibinfo  {journal} {Phys. Rev. E}\ }\textbf {\bibinfo {volume}
  {50}},\ \bibinfo {pages} {888} (\bibinfo {year} {1994})}\BibitemShut
  {NoStop}%
\bibitem [{\citenamefont {Srednicki}(1999)}]{Srednicki_1999}%
  \BibitemOpen
  \bibfield  {author} {\bibinfo {author} {\bibfnamefont {M.}~\bibnamefont
  {Srednicki}},\ }\bibfield  {title} {\bibinfo {title} {The approach to thermal
  equilibrium in quantized chaotic systems},\ }\href
  {https://doi.org/10.1088/0305-4470/32/7/007} {\bibfield  {journal} {\bibinfo
  {journal} {Journal of Physics A: Mathematical and General}\ }\textbf
  {\bibinfo {volume} {32}},\ \bibinfo {pages} {1163} (\bibinfo {year}
  {1999})}\BibitemShut {NoStop}%
\bibitem [{\citenamefont {{Haake}}(2010)}]{Haake_QuantumChaosBook}%
  \BibitemOpen
  \bibfield  {author} {\bibinfo {author} {\bibfnamefont {F.}~\bibnamefont
  {{Haake}}},\ }\href {https://doi.org/10.1007/978-3-642-05428-0} {\emph
  {\bibinfo {title} {{Quantum Signatures of Chaos}}}},\ Vol.~\bibinfo {volume}
  {54}\ (\bibinfo {year} {2010})\BibitemShut {NoStop}%
\bibitem [{\citenamefont {Cotler}\ \emph {et~al.}(2017)\citenamefont {Cotler},
  \citenamefont {Hunter-Jones}, \citenamefont {Liu},\ and\ \citenamefont
  {Yoshida}}]{Cotler_SFFChaos2017}%
  \BibitemOpen
  \bibfield  {author} {\bibinfo {author} {\bibfnamefont {J.}~\bibnamefont
  {Cotler}}, \bibinfo {author} {\bibfnamefont {N.}~\bibnamefont
  {Hunter-Jones}}, \bibinfo {author} {\bibfnamefont {J.}~\bibnamefont {Liu}},\
  and\ \bibinfo {author} {\bibfnamefont {B.}~\bibnamefont {Yoshida}},\
  }\bibfield  {title} {\bibinfo {title} {Chaos, complexity, and random
  matrices},\ }\bibfield  {journal} {\bibinfo  {journal} {Journal of High
  Energy Physics}\ }\textbf {\bibinfo {volume} {2017}},\ \href
  {https://doi.org/10.1007/jhep11(2017)048} {10.1007/jhep11(2017)048} (\bibinfo
  {year} {2017})\BibitemShut {NoStop}%
\bibitem [{\citenamefont {Gharibyan}\ \emph {et~al.}(2018)\citenamefont
  {Gharibyan}, \citenamefont {Hanada}, \citenamefont {Shenker},\ and\
  \citenamefont {Tezuka}}]{ShenkerGharibyan2018onsetofRM}%
  \BibitemOpen
  \bibfield  {author} {\bibinfo {author} {\bibfnamefont {H.}~\bibnamefont
  {Gharibyan}}, \bibinfo {author} {\bibfnamefont {M.}~\bibnamefont {Hanada}},
  \bibinfo {author} {\bibfnamefont {S.~H.}\ \bibnamefont {Shenker}},\ and\
  \bibinfo {author} {\bibfnamefont {M.}~\bibnamefont {Tezuka}},\ }\bibfield
  {title} {\bibinfo {title} {Onset of random matrix behavior in scrambling
  systems},\ }\href@noop {} {\bibfield  {journal} {\bibinfo  {journal} {Journal
  of High Energy Physics}\ }\textbf {\bibinfo {volume} {2018}},\ \bibinfo
  {pages} {124} (\bibinfo {year} {2018})}\BibitemShut {NoStop}%
\bibitem [{\citenamefont {Liu}(2018)}]{Liu_SFFChaos_PhysRevD.98.086026}%
  \BibitemOpen
  \bibfield  {author} {\bibinfo {author} {\bibfnamefont {J.}~\bibnamefont
  {Liu}},\ }\bibfield  {title} {\bibinfo {title} {Spectral form factors and
  late time quantum chaos},\ }\href
  {https://doi.org/10.1103/PhysRevD.98.086026} {\bibfield  {journal} {\bibinfo
  {journal} {Phys. Rev. D}\ }\textbf {\bibinfo {volume} {98}},\ \bibinfo
  {pages} {086026} (\bibinfo {year} {2018})}\BibitemShut {NoStop}%
\bibitem [{\citenamefont {Chen}\ and\ \citenamefont
  {Ludwig}(2018)}]{ChenLudwig_PhysRevB.98.064309}%
  \BibitemOpen
  \bibfield  {author} {\bibinfo {author} {\bibfnamefont {X.}~\bibnamefont
  {Chen}}\ and\ \bibinfo {author} {\bibfnamefont {A.~W.~W.}\ \bibnamefont
  {Ludwig}},\ }\bibfield  {title} {\bibinfo {title} {Universal spectral
  correlations in the chaotic wave function and the development of quantum
  chaos},\ }\href {https://doi.org/10.1103/PhysRevB.98.064309} {\bibfield
  {journal} {\bibinfo  {journal} {Phys. Rev. B}\ }\textbf {\bibinfo {volume}
  {98}},\ \bibinfo {pages} {064309} (\bibinfo {year} {2018})}\BibitemShut
  {NoStop}%
\bibitem [{\citenamefont {Bertini}\ \emph {et~al.}(2018)\citenamefont
  {Bertini}, \citenamefont {Kos},\ and\ \citenamefont
  {Prosen}}]{BertiniProsen_PhysRevLett.121.264101}%
  \BibitemOpen
  \bibfield  {author} {\bibinfo {author} {\bibfnamefont {B.}~\bibnamefont
  {Bertini}}, \bibinfo {author} {\bibfnamefont {P.}~\bibnamefont {Kos}},\ and\
  \bibinfo {author} {\bibfnamefont {T.~c.~v.}\ \bibnamefont {Prosen}},\
  }\bibfield  {title} {\bibinfo {title} {Exact spectral form factor in a
  minimal model of many-body quantum chaos},\ }\href
  {https://doi.org/10.1103/PhysRevLett.121.264101} {\bibfield  {journal}
  {\bibinfo  {journal} {Phys. Rev. Lett.}\ }\textbf {\bibinfo {volume} {121}},\
  \bibinfo {pages} {264101} (\bibinfo {year} {2018})}\BibitemShut {NoStop}%
\bibitem [{\citenamefont {Kos}\ \emph {et~al.}(2018)\citenamefont {Kos},
  \citenamefont {Ljubotina},\ and\ \citenamefont
  {Prosen}}]{KosLjubotinaProsen_PhysRevX.8.021062}%
  \BibitemOpen
  \bibfield  {author} {\bibinfo {author} {\bibfnamefont {P.}~\bibnamefont
  {Kos}}, \bibinfo {author} {\bibfnamefont {M.}~\bibnamefont {Ljubotina}},\
  and\ \bibinfo {author} {\bibfnamefont {T.~c.~v.}\ \bibnamefont {Prosen}},\
  }\bibfield  {title} {\bibinfo {title} {Many-body quantum chaos: Analytic
  connection to random matrix theory},\ }\href
  {https://doi.org/10.1103/PhysRevX.8.021062} {\bibfield  {journal} {\bibinfo
  {journal} {Phys. Rev. X}\ }\textbf {\bibinfo {volume} {8}},\ \bibinfo {pages}
  {021062} (\bibinfo {year} {2018})}\BibitemShut {NoStop}%
\bibitem [{\citenamefont {Nandkishore}\ and\ \citenamefont
  {Huse}(2015)}]{HuseNandkishore2015_MBLreview}%
  \BibitemOpen
  \bibfield  {author} {\bibinfo {author} {\bibfnamefont {R.}~\bibnamefont
  {Nandkishore}}\ and\ \bibinfo {author} {\bibfnamefont {D.~A.}\ \bibnamefont
  {Huse}},\ }\bibfield  {title} {\bibinfo {title} {Many-body localization and
  thermalization in quantum statistical mechanics},\ }\href@noop {} {\bibfield
  {journal} {\bibinfo  {journal} {Annu. Rev. Condens. Matter Phys.}\ }\textbf
  {\bibinfo {volume} {6}},\ \bibinfo {pages} {15} (\bibinfo {year}
  {2015})}\BibitemShut {NoStop}%
\bibitem [{\citenamefont {Abanin}\ \emph {et~al.}(2019)\citenamefont {Abanin},
  \citenamefont {Altman}, \citenamefont {Bloch},\ and\ \citenamefont
  {Serbyn}}]{abanin2019colloquium}%
  \BibitemOpen
  \bibfield  {author} {\bibinfo {author} {\bibfnamefont {D.~A.}\ \bibnamefont
  {Abanin}}, \bibinfo {author} {\bibfnamefont {E.}~\bibnamefont {Altman}},
  \bibinfo {author} {\bibfnamefont {I.}~\bibnamefont {Bloch}},\ and\ \bibinfo
  {author} {\bibfnamefont {M.}~\bibnamefont {Serbyn}},\ }\bibfield  {title}
  {\bibinfo {title} {Colloquium: Many-body localization, thermalization, and
  entanglement},\ }\href@noop {} {\bibfield  {journal} {\bibinfo  {journal}
  {Reviews of Modern Physics}\ }\textbf {\bibinfo {volume} {91}},\ \bibinfo
  {pages} {021001} (\bibinfo {year} {2019})}\BibitemShut {NoStop}%
\bibitem [{\citenamefont {Gopalakrishnan}\ and\ \citenamefont
  {Parameswaran}(2020)}]{gopalakrishnan2020dynamics}%
  \BibitemOpen
  \bibfield  {author} {\bibinfo {author} {\bibfnamefont {S.}~\bibnamefont
  {Gopalakrishnan}}\ and\ \bibinfo {author} {\bibfnamefont {S.}~\bibnamefont
  {Parameswaran}},\ }\bibfield  {title} {\bibinfo {title} {Dynamics and
  transport at the threshold of many-body localization},\ }\href@noop {}
  {\bibfield  {journal} {\bibinfo  {journal} {Physics Reports}\ } (\bibinfo
  {year} {2020})}\BibitemShut {NoStop}%
\bibitem [{\citenamefont {Deng}\ \emph {et~al.}(2017)\citenamefont {Deng},
  \citenamefont {Ganeshan}, \citenamefont {Li}, \citenamefont {Modak},
  \citenamefont {Mukerjee},\ and\ \citenamefont {Pixley}}]{deng2017many}%
  \BibitemOpen
  \bibfield  {author} {\bibinfo {author} {\bibfnamefont {D.-L.}\ \bibnamefont
  {Deng}}, \bibinfo {author} {\bibfnamefont {S.}~\bibnamefont {Ganeshan}},
  \bibinfo {author} {\bibfnamefont {X.}~\bibnamefont {Li}}, \bibinfo {author}
  {\bibfnamefont {R.}~\bibnamefont {Modak}}, \bibinfo {author} {\bibfnamefont
  {S.}~\bibnamefont {Mukerjee}},\ and\ \bibinfo {author} {\bibfnamefont
  {J.}~\bibnamefont {Pixley}},\ }\bibfield  {title} {\bibinfo {title}
  {Many-body localization in incommensurate models with a mobility edge},\
  }\href@noop {} {\bibfield  {journal} {\bibinfo  {journal} {Annalen der
  Physik}\ }\textbf {\bibinfo {volume} {529}},\ \bibinfo {pages} {1600399}
  (\bibinfo {year} {2017})}\BibitemShut {NoStop}%
\bibitem [{\citenamefont {Schreiber}\ \emph {et~al.}(2015)\citenamefont
  {Schreiber}, \citenamefont {Hodgman}, \citenamefont {Bordia}, \citenamefont
  {L{\"u}schen}, \citenamefont {Fischer}, \citenamefont {Vosk}, \citenamefont
  {Altman}, \citenamefont {Schneider},\ and\ \citenamefont
  {Bloch}}]{schreiber2015observation}%
  \BibitemOpen
  \bibfield  {author} {\bibinfo {author} {\bibfnamefont {M.}~\bibnamefont
  {Schreiber}}, \bibinfo {author} {\bibfnamefont {S.~S.}\ \bibnamefont
  {Hodgman}}, \bibinfo {author} {\bibfnamefont {P.}~\bibnamefont {Bordia}},
  \bibinfo {author} {\bibfnamefont {H.~P.}\ \bibnamefont {L{\"u}schen}},
  \bibinfo {author} {\bibfnamefont {M.~H.}\ \bibnamefont {Fischer}}, \bibinfo
  {author} {\bibfnamefont {R.}~\bibnamefont {Vosk}}, \bibinfo {author}
  {\bibfnamefont {E.}~\bibnamefont {Altman}}, \bibinfo {author} {\bibfnamefont
  {U.}~\bibnamefont {Schneider}},\ and\ \bibinfo {author} {\bibfnamefont
  {I.}~\bibnamefont {Bloch}},\ }\bibfield  {title} {\bibinfo {title}
  {Observation of many-body localization of interacting fermions in a
  quasirandom optical lattice},\ }\href@noop {} {\bibfield  {journal} {\bibinfo
   {journal} {Science}\ }\textbf {\bibinfo {volume} {349}},\ \bibinfo {pages}
  {842} (\bibinfo {year} {2015})}\BibitemShut {NoStop}%
\bibitem [{\citenamefont {Choi}\ \emph {et~al.}(2016)\citenamefont {Choi},
  \citenamefont {Hild}, \citenamefont {Zeiher}, \citenamefont {Schau{\ss}},
  \citenamefont {Rubio-Abadal}, \citenamefont {Yefsah}, \citenamefont
  {Khemani}, \citenamefont {Huse}, \citenamefont {Bloch},\ and\ \citenamefont
  {Gross}}]{choi2016exploring}%
  \BibitemOpen
  \bibfield  {author} {\bibinfo {author} {\bibfnamefont {J.-y.}\ \bibnamefont
  {Choi}}, \bibinfo {author} {\bibfnamefont {S.}~\bibnamefont {Hild}}, \bibinfo
  {author} {\bibfnamefont {J.}~\bibnamefont {Zeiher}}, \bibinfo {author}
  {\bibfnamefont {P.}~\bibnamefont {Schau{\ss}}}, \bibinfo {author}
  {\bibfnamefont {A.}~\bibnamefont {Rubio-Abadal}}, \bibinfo {author}
  {\bibfnamefont {T.}~\bibnamefont {Yefsah}}, \bibinfo {author} {\bibfnamefont
  {V.}~\bibnamefont {Khemani}}, \bibinfo {author} {\bibfnamefont {D.~A.}\
  \bibnamefont {Huse}}, \bibinfo {author} {\bibfnamefont {I.}~\bibnamefont
  {Bloch}},\ and\ \bibinfo {author} {\bibfnamefont {C.}~\bibnamefont {Gross}},\
  }\bibfield  {title} {\bibinfo {title} {Exploring the many-body localization
  transition in two dimensions},\ }\href@noop {} {\bibfield  {journal}
  {\bibinfo  {journal} {Science}\ }\textbf {\bibinfo {volume} {352}},\ \bibinfo
  {pages} {1547} (\bibinfo {year} {2016})}\BibitemShut {NoStop}%
\bibitem [{\citenamefont {Lukin}\ \emph {et~al.}(2019)\citenamefont {Lukin},
  \citenamefont {Rispoli}, \citenamefont {Schittko}, \citenamefont {Tai},
  \citenamefont {Kaufman}, \citenamefont {Choi}, \citenamefont {Khemani},
  \citenamefont {L{\'e}onard},\ and\ \citenamefont
  {Greiner}}]{lukin2019probing}%
  \BibitemOpen
  \bibfield  {author} {\bibinfo {author} {\bibfnamefont {A.}~\bibnamefont
  {Lukin}}, \bibinfo {author} {\bibfnamefont {M.}~\bibnamefont {Rispoli}},
  \bibinfo {author} {\bibfnamefont {R.}~\bibnamefont {Schittko}}, \bibinfo
  {author} {\bibfnamefont {M.~E.}\ \bibnamefont {Tai}}, \bibinfo {author}
  {\bibfnamefont {A.~M.}\ \bibnamefont {Kaufman}}, \bibinfo {author}
  {\bibfnamefont {S.}~\bibnamefont {Choi}}, \bibinfo {author} {\bibfnamefont
  {V.}~\bibnamefont {Khemani}}, \bibinfo {author} {\bibfnamefont
  {J.}~\bibnamefont {L{\'e}onard}},\ and\ \bibinfo {author} {\bibfnamefont
  {M.}~\bibnamefont {Greiner}},\ }\bibfield  {title} {\bibinfo {title} {Probing
  entanglement in a many-body--localized system},\ }\href@noop {} {\bibfield
  {journal} {\bibinfo  {journal} {Science}\ }\textbf {\bibinfo {volume}
  {364}},\ \bibinfo {pages} {256} (\bibinfo {year} {2019})}\BibitemShut
  {NoStop}%
\bibitem [{\citenamefont {Smith}\ \emph {et~al.}(2016)\citenamefont {Smith},
  \citenamefont {Lee}, \citenamefont {Richerme}, \citenamefont {Neyenhuis},
  \citenamefont {Hess}, \citenamefont {Hauke}, \citenamefont {Heyl},
  \citenamefont {Huse},\ and\ \citenamefont {Monroe}}]{Smith_2016_MBLExpt}%
  \BibitemOpen
  \bibfield  {author} {\bibinfo {author} {\bibfnamefont {J.}~\bibnamefont
  {Smith}}, \bibinfo {author} {\bibfnamefont {A.}~\bibnamefont {Lee}}, \bibinfo
  {author} {\bibfnamefont {P.}~\bibnamefont {Richerme}}, \bibinfo {author}
  {\bibfnamefont {B.}~\bibnamefont {Neyenhuis}}, \bibinfo {author}
  {\bibfnamefont {P.~W.}\ \bibnamefont {Hess}}, \bibinfo {author}
  {\bibfnamefont {P.}~\bibnamefont {Hauke}}, \bibinfo {author} {\bibfnamefont
  {M.}~\bibnamefont {Heyl}}, \bibinfo {author} {\bibfnamefont {D.~A.}\
  \bibnamefont {Huse}},\ and\ \bibinfo {author} {\bibfnamefont
  {C.}~\bibnamefont {Monroe}},\ }\bibfield  {title} {\bibinfo {title}
  {Many-body localization in a quantum simulator with programmable random
  disorder},\ }\href {https://doi.org/10.1038/nphys3783} {\bibfield  {journal}
  {\bibinfo  {journal} {Nature Physics}\ }\textbf {\bibinfo {volume} {12}},\
  \bibinfo {pages} {907–911} (\bibinfo {year} {2016})}\BibitemShut {NoStop}%
\bibitem [{\citenamefont {Roushan}\ \emph {et~al.}(2017)\citenamefont
  {Roushan}, \citenamefont {Neill}, \citenamefont {Tangpanitanon},
  \citenamefont {Bastidas}, \citenamefont {Megrant}, \citenamefont {Barends},
  \citenamefont {Chen}, \citenamefont {Chen}, \citenamefont {Chiaro},
  \citenamefont {Dunsworth} \emph {et~al.}}]{roushan2017spectroscopic}%
  \BibitemOpen
  \bibfield  {author} {\bibinfo {author} {\bibfnamefont {P.}~\bibnamefont
  {Roushan}}, \bibinfo {author} {\bibfnamefont {C.}~\bibnamefont {Neill}},
  \bibinfo {author} {\bibfnamefont {J.}~\bibnamefont {Tangpanitanon}}, \bibinfo
  {author} {\bibfnamefont {V.}~\bibnamefont {Bastidas}}, \bibinfo {author}
  {\bibfnamefont {A.}~\bibnamefont {Megrant}}, \bibinfo {author} {\bibfnamefont
  {R.}~\bibnamefont {Barends}}, \bibinfo {author} {\bibfnamefont
  {Y.}~\bibnamefont {Chen}}, \bibinfo {author} {\bibfnamefont {Z.}~\bibnamefont
  {Chen}}, \bibinfo {author} {\bibfnamefont {B.}~\bibnamefont {Chiaro}},
  \bibinfo {author} {\bibfnamefont {A.}~\bibnamefont {Dunsworth}}, \emph
  {et~al.},\ }\bibfield  {title} {\bibinfo {title} {Spectroscopic signatures of
  localization with interacting photons in superconducting qubits},\
  }\href@noop {} {\bibfield  {journal} {\bibinfo  {journal} {Science}\ }\textbf
  {\bibinfo {volume} {358}},\ \bibinfo {pages} {1175} (\bibinfo {year}
  {2017})}\BibitemShut {NoStop}%
\bibitem [{\citenamefont {Xu}\ \emph {et~al.}(2018)\citenamefont {Xu},
  \citenamefont {Chen}, \citenamefont {Zeng}, \citenamefont {Zhang},
  \citenamefont {Song}, \citenamefont {Liu}, \citenamefont {Guo}, \citenamefont
  {Zhang}, \citenamefont {Xu}, \citenamefont {Deng} \emph
  {et~al.}}]{xu2018emulating}%
  \BibitemOpen
  \bibfield  {author} {\bibinfo {author} {\bibfnamefont {K.}~\bibnamefont
  {Xu}}, \bibinfo {author} {\bibfnamefont {J.-J.}\ \bibnamefont {Chen}},
  \bibinfo {author} {\bibfnamefont {Y.}~\bibnamefont {Zeng}}, \bibinfo {author}
  {\bibfnamefont {Y.-R.}\ \bibnamefont {Zhang}}, \bibinfo {author}
  {\bibfnamefont {C.}~\bibnamefont {Song}}, \bibinfo {author} {\bibfnamefont
  {W.}~\bibnamefont {Liu}}, \bibinfo {author} {\bibfnamefont {Q.}~\bibnamefont
  {Guo}}, \bibinfo {author} {\bibfnamefont {P.}~\bibnamefont {Zhang}}, \bibinfo
  {author} {\bibfnamefont {D.}~\bibnamefont {Xu}}, \bibinfo {author}
  {\bibfnamefont {H.}~\bibnamefont {Deng}}, \emph {et~al.},\ }\bibfield
  {title} {\bibinfo {title} {Emulating many-body localization with a
  superconducting quantum processor},\ }\href@noop {} {\bibfield  {journal}
  {\bibinfo  {journal} {Physical review letters}\ }\textbf {\bibinfo {volume}
  {120}},\ \bibinfo {pages} {050507} (\bibinfo {year} {2018})}\BibitemShut
  {NoStop}%
\bibitem [{\citenamefont {Wei}\ \emph {et~al.}(2018)\citenamefont {Wei},
  \citenamefont {Ramanathan},\ and\ \citenamefont
  {Cappellaro}}]{wei2018exploring}%
  \BibitemOpen
  \bibfield  {author} {\bibinfo {author} {\bibfnamefont {K.~X.}\ \bibnamefont
  {Wei}}, \bibinfo {author} {\bibfnamefont {C.}~\bibnamefont {Ramanathan}},\
  and\ \bibinfo {author} {\bibfnamefont {P.}~\bibnamefont {Cappellaro}},\
  }\bibfield  {title} {\bibinfo {title} {Exploring localization in nuclear spin
  chains},\ }\href@noop {} {\bibfield  {journal} {\bibinfo  {journal} {Physical
  review letters}\ }\textbf {\bibinfo {volume} {120}},\ \bibinfo {pages}
  {070501} (\bibinfo {year} {2018})}\BibitemShut {NoStop}%
\bibitem [{\citenamefont {De~Roeck}\ and\ \citenamefont
  {Huveneers}(2017)}]{DeRoekHuveneers_bubble_PhysRevB.95.155129}%
  \BibitemOpen
  \bibfield  {author} {\bibinfo {author} {\bibfnamefont {W.}~\bibnamefont
  {De~Roeck}}\ and\ \bibinfo {author} {\bibfnamefont {F.~m.~c.}\ \bibnamefont
  {Huveneers}},\ }\bibfield  {title} {\bibinfo {title} {Stability and
  instability towards delocalization in many-body localization systems},\
  }\href {https://doi.org/10.1103/PhysRevB.95.155129} {\bibfield  {journal}
  {\bibinfo  {journal} {Phys. Rev. B}\ }\textbf {\bibinfo {volume} {95}},\
  \bibinfo {pages} {155129} (\bibinfo {year} {2017})}\BibitemShut {NoStop}%
\bibitem [{\citenamefont {Potter}\ and\ \citenamefont
  {Vasseur}(2016)}]{PotterVasseur_PhysRevB.94.224206}%
  \BibitemOpen
  \bibfield  {author} {\bibinfo {author} {\bibfnamefont {A.~C.}\ \bibnamefont
  {Potter}}\ and\ \bibinfo {author} {\bibfnamefont {R.}~\bibnamefont
  {Vasseur}},\ }\bibfield  {title} {\bibinfo {title} {Symmetry constraints on
  many-body localization},\ }\href {https://doi.org/10.1103/PhysRevB.94.224206}
  {\bibfield  {journal} {\bibinfo  {journal} {Phys. Rev. B}\ }\textbf {\bibinfo
  {volume} {94}},\ \bibinfo {pages} {224206} (\bibinfo {year}
  {2016})}\BibitemShut {NoStop}%
\bibitem [{\citenamefont {\ifmmode~\check{S}\else \v{S}\fi{}untajs}\ \emph
  {et~al.}(2020{\natexlab{a}})\citenamefont {\ifmmode~\check{S}\else
  \v{S}\fi{}untajs}, \citenamefont {Bon\ifmmode~\check{c}\else \v{c}\fi{}a},
  \citenamefont {Prosen},\ and\ \citenamefont
  {Vidmar}}]{Prosen_2019arXiv190506345S}%
  \BibitemOpen
  \bibfield  {author} {\bibinfo {author} {\bibfnamefont {J.}~\bibnamefont
  {\ifmmode~\check{S}\else \v{S}\fi{}untajs}}, \bibinfo {author} {\bibfnamefont
  {J.}~\bibnamefont {Bon\ifmmode~\check{c}\else \v{c}\fi{}a}}, \bibinfo
  {author} {\bibfnamefont {T.~c.~v.}\ \bibnamefont {Prosen}},\ and\ \bibinfo
  {author} {\bibfnamefont {L.}~\bibnamefont {Vidmar}},\ }\bibfield  {title}
  {\bibinfo {title} {Quantum chaos challenges many-body localization},\ }\href
  {https://doi.org/10.1103/PhysRevE.102.062144} {\bibfield  {journal} {\bibinfo
   {journal} {Phys. Rev. E}\ }\textbf {\bibinfo {volume} {102}},\ \bibinfo
  {pages} {062144} (\bibinfo {year} {2020}{\natexlab{a}})}\BibitemShut
  {NoStop}%
\bibitem [{\citenamefont {Abanin}\ \emph {et~al.}(2021)\citenamefont {Abanin},
  \citenamefont {Bardarson}, \citenamefont {{De Tomasi}}, \citenamefont
  {Gopalakrishnan}, \citenamefont {Khemani}, \citenamefont {Parameswaran},
  \citenamefont {Pollmann}, \citenamefont {Potter}, \citenamefont {Serbyn},\
  and\ \citenamefont {Vasseur}}]{Abaninetal_ReplyToProsen_2019arXiv191104501A}%
  \BibitemOpen
  \bibfield  {author} {\bibinfo {author} {\bibfnamefont {D.}~\bibnamefont
  {Abanin}}, \bibinfo {author} {\bibfnamefont {J.}~\bibnamefont {Bardarson}},
  \bibinfo {author} {\bibfnamefont {G.}~\bibnamefont {{De Tomasi}}}, \bibinfo
  {author} {\bibfnamefont {S.}~\bibnamefont {Gopalakrishnan}}, \bibinfo
  {author} {\bibfnamefont {V.}~\bibnamefont {Khemani}}, \bibinfo {author}
  {\bibfnamefont {S.}~\bibnamefont {Parameswaran}}, \bibinfo {author}
  {\bibfnamefont {F.}~\bibnamefont {Pollmann}}, \bibinfo {author}
  {\bibfnamefont {A.}~\bibnamefont {Potter}}, \bibinfo {author} {\bibfnamefont
  {M.}~\bibnamefont {Serbyn}},\ and\ \bibinfo {author} {\bibfnamefont
  {R.}~\bibnamefont {Vasseur}},\ }\bibfield  {title} {\bibinfo {title}
  {Distinguishing localization from chaos: Challenges in finite-size systems},\
  }\href {https://doi.org/https://doi.org/10.1016/j.aop.2021.168415} {\bibfield
   {journal} {\bibinfo  {journal} {Annals of Physics}\ ,\ \bibinfo {pages}
  {168415}} (\bibinfo {year} {2021})}\BibitemShut {NoStop}%
\bibitem [{\citenamefont {Sierant}\ \emph {et~al.}(2020)\citenamefont
  {Sierant}, \citenamefont {Delande},\ and\ \citenamefont
  {Zakrzewski}}]{sierant2020thouless}%
  \BibitemOpen
  \bibfield  {author} {\bibinfo {author} {\bibfnamefont {P.}~\bibnamefont
  {Sierant}}, \bibinfo {author} {\bibfnamefont {D.}~\bibnamefont {Delande}},\
  and\ \bibinfo {author} {\bibfnamefont {J.}~\bibnamefont {Zakrzewski}},\
  }\bibfield  {title} {\bibinfo {title} {Thouless time analysis of anderson and
  many-body localization transitions},\ }\href@noop {} {\bibfield  {journal}
  {\bibinfo  {journal} {Physical Review Letters}\ }\textbf {\bibinfo {volume}
  {124}},\ \bibinfo {pages} {186601} (\bibinfo {year} {2020})}\BibitemShut
  {NoStop}%
\bibitem [{\citenamefont {Panda}\ \emph {et~al.}(2020)\citenamefont {Panda},
  \citenamefont {Scardicchio}, \citenamefont {Schulz}, \citenamefont {Taylor},\
  and\ \citenamefont {{\v{Z}}nidari{\v{c}}}}]{panda2020can}%
  \BibitemOpen
  \bibfield  {author} {\bibinfo {author} {\bibfnamefont {R.~K.}\ \bibnamefont
  {Panda}}, \bibinfo {author} {\bibfnamefont {A.}~\bibnamefont {Scardicchio}},
  \bibinfo {author} {\bibfnamefont {M.}~\bibnamefont {Schulz}}, \bibinfo
  {author} {\bibfnamefont {S.~R.}\ \bibnamefont {Taylor}},\ and\ \bibinfo
  {author} {\bibfnamefont {M.}~\bibnamefont {{\v{Z}}nidari{\v{c}}}},\
  }\bibfield  {title} {\bibinfo {title} {Can we study the many-body
  localisation transition?},\ }\href@noop {} {\bibfield  {journal} {\bibinfo
  {journal} {EPL (Europhysics Letters)}\ }\textbf {\bibinfo {volume} {128}},\
  \bibinfo {pages} {67003} (\bibinfo {year} {2020})}\BibitemShut {NoStop}%
\bibitem [{\citenamefont {\ifmmode~\check{S}\else \v{S}\fi{}untajs}\ \emph
  {et~al.}(2020{\natexlab{b}})\citenamefont {\ifmmode~\check{S}\else
  \v{S}\fi{}untajs}, \citenamefont {Bon\ifmmode~\check{c}\else \v{c}\fi{}a},
  \citenamefont {Prosen},\ and\ \citenamefont
  {Vidmar}}]{vsuntajs2020ergodicity}%
  \BibitemOpen
  \bibfield  {author} {\bibinfo {author} {\bibfnamefont {J.}~\bibnamefont
  {\ifmmode~\check{S}\else \v{S}\fi{}untajs}}, \bibinfo {author} {\bibfnamefont
  {J.}~\bibnamefont {Bon\ifmmode~\check{c}\else \v{c}\fi{}a}}, \bibinfo
  {author} {\bibfnamefont {T.~c.~v.}\ \bibnamefont {Prosen}},\ and\ \bibinfo
  {author} {\bibfnamefont {L.}~\bibnamefont {Vidmar}},\ }\bibfield  {title}
  {\bibinfo {title} {Ergodicity breaking transition in finite disordered spin
  chains},\ }\href {https://doi.org/10.1103/PhysRevB.102.064207} {\bibfield
  {journal} {\bibinfo  {journal} {Phys. Rev. B}\ }\textbf {\bibinfo {volume}
  {102}},\ \bibinfo {pages} {064207} (\bibinfo {year}
  {2020}{\natexlab{b}})}\BibitemShut {NoStop}%
\bibitem [{\citenamefont {Goremykina}\ \emph {et~al.}(2019)\citenamefont
  {Goremykina}, \citenamefont {Vasseur},\ and\ \citenamefont
  {Serbyn}}]{goremykina2019analytically}%
  \BibitemOpen
  \bibfield  {author} {\bibinfo {author} {\bibfnamefont {A.}~\bibnamefont
  {Goremykina}}, \bibinfo {author} {\bibfnamefont {R.}~\bibnamefont
  {Vasseur}},\ and\ \bibinfo {author} {\bibfnamefont {M.}~\bibnamefont
  {Serbyn}},\ }\bibfield  {title} {\bibinfo {title} {Analytically solvable
  renormalization group for the many-body localization transition},\
  }\href@noop {} {\bibfield  {journal} {\bibinfo  {journal} {Physical review
  letters}\ }\textbf {\bibinfo {volume} {122}},\ \bibinfo {pages} {040601}
  (\bibinfo {year} {2019})}\BibitemShut {NoStop}%
\bibitem [{\citenamefont {Dumitrescu}\ \emph {et~al.}(2019)\citenamefont
  {Dumitrescu}, \citenamefont {Goremykina}, \citenamefont {Parameswaran},
  \citenamefont {Serbyn},\ and\ \citenamefont
  {Vasseur}}]{dumitrescu2019kosterlitz}%
  \BibitemOpen
  \bibfield  {author} {\bibinfo {author} {\bibfnamefont {P.~T.}\ \bibnamefont
  {Dumitrescu}}, \bibinfo {author} {\bibfnamefont {A.}~\bibnamefont
  {Goremykina}}, \bibinfo {author} {\bibfnamefont {S.~A.}\ \bibnamefont
  {Parameswaran}}, \bibinfo {author} {\bibfnamefont {M.}~\bibnamefont
  {Serbyn}},\ and\ \bibinfo {author} {\bibfnamefont {R.}~\bibnamefont
  {Vasseur}},\ }\bibfield  {title} {\bibinfo {title} {Kosterlitz-thouless
  scaling at many-body localization phase transitions},\ }\href@noop {}
  {\bibfield  {journal} {\bibinfo  {journal} {Physical Review B}\ }\textbf
  {\bibinfo {volume} {99}},\ \bibinfo {pages} {094205} (\bibinfo {year}
  {2019})}\BibitemShut {NoStop}%
\bibitem [{\citenamefont {Morningstar}\ \emph {et~al.}(2020)\citenamefont
  {Morningstar}, \citenamefont {Huse},\ and\ \citenamefont
  {Imbrie}}]{morningstar2020many}%
  \BibitemOpen
  \bibfield  {author} {\bibinfo {author} {\bibfnamefont {A.}~\bibnamefont
  {Morningstar}}, \bibinfo {author} {\bibfnamefont {D.~A.}\ \bibnamefont
  {Huse}},\ and\ \bibinfo {author} {\bibfnamefont {J.~Z.}\ \bibnamefont
  {Imbrie}},\ }\bibfield  {title} {\bibinfo {title} {Many-body localization
  near the critical point},\ }\href
  {https://doi.org/10.1103/PhysRevB.102.125134} {\bibfield  {journal} {\bibinfo
   {journal} {Phys. Rev. B}\ }\textbf {\bibinfo {volume} {102}},\ \bibinfo
  {pages} {125134} (\bibinfo {year} {2020})}\BibitemShut {NoStop}%
\bibitem [{\citenamefont {Chan}\ \emph {et~al.}(2018)\citenamefont {Chan},
  \citenamefont {De~Luca},\ and\ \citenamefont
  {Chalker}}]{ChanLucaChalker_PhysRevLett.121.060601}%
  \BibitemOpen
  \bibfield  {author} {\bibinfo {author} {\bibfnamefont {A.}~\bibnamefont
  {Chan}}, \bibinfo {author} {\bibfnamefont {A.}~\bibnamefont {De~Luca}},\ and\
  \bibinfo {author} {\bibfnamefont {J.~T.}\ \bibnamefont {Chalker}},\
  }\bibfield  {title} {\bibinfo {title} {Spectral statistics in spatially
  extended chaotic quantum many-body systems},\ }\href
  {https://doi.org/10.1103/PhysRevLett.121.060601} {\bibfield  {journal}
  {\bibinfo  {journal} {Phys. Rev. Lett.}\ }\textbf {\bibinfo {volume} {121}},\
  \bibinfo {pages} {060601} (\bibinfo {year} {2018})}\BibitemShut {NoStop}%
\bibitem [{\citenamefont {Oganesyan}\ and\ \citenamefont
  {Huse}(2007)}]{OganesyanHuse_2007_PhysRevB.75.155111}%
  \BibitemOpen
  \bibfield  {author} {\bibinfo {author} {\bibfnamefont {V.}~\bibnamefont
  {Oganesyan}}\ and\ \bibinfo {author} {\bibfnamefont {D.~A.}\ \bibnamefont
  {Huse}},\ }\bibfield  {title} {\bibinfo {title} {Localization of interacting
  fermions at high temperature},\ }\href
  {https://doi.org/10.1103/PhysRevB.75.155111} {\bibfield  {journal} {\bibinfo
  {journal} {Phys. Rev. B}\ }\textbf {\bibinfo {volume} {75}},\ \bibinfo
  {pages} {155111} (\bibinfo {year} {2007})}\BibitemShut {NoStop}%
\bibitem [{sup()}]{supp}%
  \BibitemOpen
  \href@noop {} {\bibinfo  {journal} {Supplemental Material}\ }\BibitemShut
  {NoStop}%
\bibitem [{\citenamefont {Riser}\ \emph {et~al.}(2020)\citenamefont {Riser},
  \citenamefont {Osipov},\ and\ \citenamefont
  {Kanzieper}}]{riser2020nonperturbative}%
  \BibitemOpen
\bibfield  {journal} {  }\bibfield  {author} {\bibinfo {author} {\bibfnamefont
  {R.}~\bibnamefont {Riser}}, \bibinfo {author} {\bibfnamefont {V.~A.}\
  \bibnamefont {Osipov}},\ and\ \bibinfo {author} {\bibfnamefont
  {E.}~\bibnamefont {Kanzieper}},\ }\bibfield  {title} {\bibinfo {title}
  {Nonperturbative theory of power spectrum in complex systems},\ }\href@noop
  {} {\bibfield  {journal} {\bibinfo  {journal} {Annals of Physics}\ }\textbf
  {\bibinfo {volume} {413}},\ \bibinfo {pages} {168065} (\bibinfo {year}
  {2020})}\BibitemShut {NoStop}%
\bibitem [{\citenamefont {Serbyn}\ \emph {et~al.}(2013)\citenamefont {Serbyn},
  \citenamefont {Papi\ifmmode~\acute{c}\else \'{c}\fi{}},\ and\ \citenamefont
  {Abanin}}]{SerbinPapicAbanin_lbit_PhysRevLett.111.127201}%
  \BibitemOpen
  \bibfield  {author} {\bibinfo {author} {\bibfnamefont {M.}~\bibnamefont
  {Serbyn}}, \bibinfo {author} {\bibfnamefont {Z.}~\bibnamefont
  {Papi\ifmmode~\acute{c}\else \'{c}\fi{}}},\ and\ \bibinfo {author}
  {\bibfnamefont {D.~A.}\ \bibnamefont {Abanin}},\ }\bibfield  {title}
  {\bibinfo {title} {Local conservation laws and the structure of the many-body
  localized states},\ }\href {https://doi.org/10.1103/PhysRevLett.111.127201}
  {\bibfield  {journal} {\bibinfo  {journal} {Phys. Rev. Lett.}\ }\textbf
  {\bibinfo {volume} {111}},\ \bibinfo {pages} {127201} (\bibinfo {year}
  {2013})}\BibitemShut {NoStop}%
\bibitem [{\citenamefont {Huse}\ \emph {et~al.}(2014)\citenamefont {Huse},
  \citenamefont {Nandkishore},\ and\ \citenamefont
  {Oganesyan}}]{HuseOganesyan_lbit_PhysRevB.90.174202}%
  \BibitemOpen
  \bibfield  {author} {\bibinfo {author} {\bibfnamefont {D.~A.}\ \bibnamefont
  {Huse}}, \bibinfo {author} {\bibfnamefont {R.}~\bibnamefont {Nandkishore}},\
  and\ \bibinfo {author} {\bibfnamefont {V.}~\bibnamefont {Oganesyan}},\
  }\bibfield  {title} {\bibinfo {title} {Phenomenology of fully
  many-body-localized systems},\ }\href
  {https://doi.org/10.1103/PhysRevB.90.174202} {\bibfield  {journal} {\bibinfo
  {journal} {Phys. Rev. B}\ }\textbf {\bibinfo {volume} {90}},\ \bibinfo
  {pages} {174202} (\bibinfo {year} {2014})}\BibitemShut {NoStop}%
\bibitem [{\citenamefont {Berry}\ and\ \citenamefont
  {Tabor}(1977)}]{BerryTabor1977_level}%
  \BibitemOpen
  \bibfield  {author} {\bibinfo {author} {\bibfnamefont {M.~V.}\ \bibnamefont
  {Berry}}\ and\ \bibinfo {author} {\bibfnamefont {M.}~\bibnamefont {Tabor}},\
  }\bibfield  {title} {\bibinfo {title} {Level clustering in the regular
  spectrum},\ }\href@noop {} {\bibfield  {journal} {\bibinfo  {journal}
  {Proceedings of the Royal Society of London. A. Mathematical and Physical
  Sciences}\ }\textbf {\bibinfo {volume} {356}},\ \bibinfo {pages} {375}
  (\bibinfo {year} {1977})}\BibitemShut {NoStop}%
\bibitem [{\citenamefont {Marklof}(2001)}]{BerryTabor_conjecture_Jens}%
  \BibitemOpen
  \bibfield  {author} {\bibinfo {author} {\bibfnamefont {J.}~\bibnamefont
  {Marklof}},\ }\bibfield  {title} {\bibinfo {title} {The berry-tabor
  conjecture},\ }in\ \href@noop {} {\emph {\bibinfo {booktitle} {European
  Congress of Mathematics}}},\ \bibinfo {editor} {edited by\ \bibinfo {editor}
  {\bibfnamefont {C.}~\bibnamefont {Casacuberta}}, \bibinfo {editor}
  {\bibfnamefont {R.~M.}\ \bibnamefont {Mir{\'o}-Roig}}, \bibinfo {editor}
  {\bibfnamefont {J.}~\bibnamefont {Verdera}},\ and\ \bibinfo {editor}
  {\bibfnamefont {S.}~\bibnamefont {Xamb{\'o}-Descamps}}}\ (\bibinfo
  {publisher} {Birkh{\"a}user Basel},\ \bibinfo {address} {Basel},\ \bibinfo
  {year} {2001})\ pp.\ \bibinfo {pages} {421--427}\BibitemShut {NoStop}%
\bibitem [{\citenamefont {Khemani}\ \emph
  {et~al.}(2017{\natexlab{a}})\citenamefont {Khemani}, \citenamefont {Sheng},\
  and\ \citenamefont
  {Huse}}]{KhemaniShengHuse_CriricalMBL_PhysRevLett.119.075702}%
  \BibitemOpen
  \bibfield  {author} {\bibinfo {author} {\bibfnamefont {V.}~\bibnamefont
  {Khemani}}, \bibinfo {author} {\bibfnamefont {D.~N.}\ \bibnamefont {Sheng}},\
  and\ \bibinfo {author} {\bibfnamefont {D.~A.}\ \bibnamefont {Huse}},\
  }\bibfield  {title} {\bibinfo {title} {Two universality classes for the
  many-body localization transition},\ }\href
  {https://doi.org/10.1103/PhysRevLett.119.075702} {\bibfield  {journal}
  {\bibinfo  {journal} {Phys. Rev. Lett.}\ }\textbf {\bibinfo {volume} {119}},\
  \bibinfo {pages} {075702} (\bibinfo {year} {2017}{\natexlab{a}})}\BibitemShut
  {NoStop}%
\bibitem [{\citenamefont {Khemani}\ \emph
  {et~al.}(2017{\natexlab{b}})\citenamefont {Khemani}, \citenamefont {Lim},
  \citenamefont {Sheng},\ and\ \citenamefont {Huse}}]{khemani2017critical}%
  \BibitemOpen
  \bibfield  {author} {\bibinfo {author} {\bibfnamefont {V.}~\bibnamefont
  {Khemani}}, \bibinfo {author} {\bibfnamefont {S.-P.}\ \bibnamefont {Lim}},
  \bibinfo {author} {\bibfnamefont {D.}~\bibnamefont {Sheng}},\ and\ \bibinfo
  {author} {\bibfnamefont {D.~A.}\ \bibnamefont {Huse}},\ }\bibfield  {title}
  {\bibinfo {title} {Critical properties of the many-body localization
  transition},\ }\href@noop {} {\bibfield  {journal} {\bibinfo  {journal}
  {Physical Review X}\ }\textbf {\bibinfo {volume} {7}},\ \bibinfo {pages}
  {021013} (\bibinfo {year} {2017}{\natexlab{b}})}\BibitemShut {NoStop}%
\bibitem [{\citenamefont
  {Anderson}(1958)}]{AndersonLocalization_PhysRev.109.1492}%
  \BibitemOpen
  \bibfield  {author} {\bibinfo {author} {\bibfnamefont {P.~W.}\ \bibnamefont
  {Anderson}},\ }\bibfield  {title} {\bibinfo {title} {Absence of diffusion in
  certain random lattices},\ }\href {https://doi.org/10.1103/PhysRev.109.1492}
  {\bibfield  {journal} {\bibinfo  {journal} {Phys. Rev.}\ }\textbf {\bibinfo
  {volume} {109}},\ \bibinfo {pages} {1492} (\bibinfo {year}
  {1958})}\BibitemShut {NoStop}%
\bibitem [{\citenamefont {Mehta}(2004)}]{mehta2004random}%
  \BibitemOpen
  \bibfield  {author} {\bibinfo {author} {\bibfnamefont {M.}~\bibnamefont
  {Mehta}},\ }\href {https://books.google.co.in/books?id=Kp3Nx03\_gMwC} {\emph
  {\bibinfo {title} {Random Matrices}}},\ ISSN\ (\bibinfo  {publisher}
  {Elsevier Science},\ \bibinfo {year} {2004})\BibitemShut {NoStop}%
\bibitem [{\citenamefont {Casati}\ \emph {et~al.}(1985)\citenamefont {Casati},
  \citenamefont {Chirikov},\ and\ \citenamefont
  {Guarneri}}]{Casati_billiards_PhysRevLett.54.1350}%
  \BibitemOpen
  \bibfield  {author} {\bibinfo {author} {\bibfnamefont {G.}~\bibnamefont
  {Casati}}, \bibinfo {author} {\bibfnamefont {B.~V.}\ \bibnamefont
  {Chirikov}},\ and\ \bibinfo {author} {\bibfnamefont {I.}~\bibnamefont
  {Guarneri}},\ }\bibfield  {title} {\bibinfo {title} {Energy-level statistics
  of integrable quantum systems},\ }\href
  {https://doi.org/10.1103/PhysRevLett.54.1350} {\bibfield  {journal} {\bibinfo
   {journal} {Phys. Rev. Lett.}\ }\textbf {\bibinfo {volume} {54}},\ \bibinfo
  {pages} {1350} (\bibinfo {year} {1985})}\BibitemShut {NoStop}%
\bibitem [{\citenamefont {Marklof}(1998)}]{marklof1998spectral_billiards}%
  \BibitemOpen
  \bibfield  {author} {\bibinfo {author} {\bibfnamefont {J.}~\bibnamefont
  {Marklof}},\ }\bibfield  {title} {\bibinfo {title} {Spectral form factors of
  rectangle billiards},\ }\href@noop {} {\bibfield  {journal} {\bibinfo
  {journal} {Communications in mathematical physics}\ }\textbf {\bibinfo
  {volume} {199}},\ \bibinfo {pages} {169} (\bibinfo {year}
  {1998})}\BibitemShut {NoStop}%
\bibitem [{\citenamefont {Serbyn}\ and\ \citenamefont
  {Moore}(2016)}]{SerbynMoore_PhysRevB.93.041424}%
  \BibitemOpen
  \bibfield  {author} {\bibinfo {author} {\bibfnamefont {M.}~\bibnamefont
  {Serbyn}}\ and\ \bibinfo {author} {\bibfnamefont {J.~E.}\ \bibnamefont
  {Moore}},\ }\bibfield  {title} {\bibinfo {title} {Spectral statistics across
  the many-body localization transition},\ }\href
  {https://doi.org/10.1103/PhysRevB.93.041424} {\bibfield  {journal} {\bibinfo
  {journal} {Phys. Rev. B}\ }\textbf {\bibinfo {volume} {93}},\ \bibinfo
  {pages} {041424} (\bibinfo {year} {2016})}\BibitemShut {NoStop}%
\bibitem [{\citenamefont {Sierant}\ and\ \citenamefont
  {Zakrzewski}(2019)}]{LevelstatistticsMBL_PhysRevB.99.104205}%
  \BibitemOpen
  \bibfield  {author} {\bibinfo {author} {\bibfnamefont {P.}~\bibnamefont
  {Sierant}}\ and\ \bibinfo {author} {\bibfnamefont {J.}~\bibnamefont
  {Zakrzewski}},\ }\bibfield  {title} {\bibinfo {title} {Level statistics
  across the many-body localization transition},\ }\href
  {https://doi.org/10.1103/PhysRevB.99.104205} {\bibfield  {journal} {\bibinfo
  {journal} {Phys. Rev. B}\ }\textbf {\bibinfo {volume} {99}},\ \bibinfo
  {pages} {104205} (\bibinfo {year} {2019})}\BibitemShut {NoStop}%
\bibitem [{\citenamefont {Buijsman}\ \emph {et~al.}(2019)\citenamefont
  {Buijsman}, \citenamefont {Cheianov},\ and\ \citenamefont
  {Gritsev}}]{RMT_MBL_PhysRevLett.122.180601}%
  \BibitemOpen
  \bibfield  {author} {\bibinfo {author} {\bibfnamefont {W.}~\bibnamefont
  {Buijsman}}, \bibinfo {author} {\bibfnamefont {V.}~\bibnamefont {Cheianov}},\
  and\ \bibinfo {author} {\bibfnamefont {V.}~\bibnamefont {Gritsev}},\
  }\bibfield  {title} {\bibinfo {title} {Random matrix ensemble for the level
  statistics of many-body localization},\ }\href
  {https://doi.org/10.1103/PhysRevLett.122.180601} {\bibfield  {journal}
  {\bibinfo  {journal} {Phys. Rev. Lett.}\ }\textbf {\bibinfo {volume} {122}},\
  \bibinfo {pages} {180601} (\bibinfo {year} {2019})}\BibitemShut {NoStop}%
\bibitem [{\citenamefont {Sierant}\ and\ \citenamefont
  {Zakrzewski}(2020)}]{RMT_MBL_PhysRevB.101.104201}%
  \BibitemOpen
  \bibfield  {author} {\bibinfo {author} {\bibfnamefont {P.}~\bibnamefont
  {Sierant}}\ and\ \bibinfo {author} {\bibfnamefont {J.}~\bibnamefont
  {Zakrzewski}},\ }\bibfield  {title} {\bibinfo {title} {Model of level
  statistics for disordered interacting quantum many-body systems},\ }\href
  {https://doi.org/10.1103/PhysRevB.101.104201} {\bibfield  {journal} {\bibinfo
   {journal} {Phys. Rev. B}\ }\textbf {\bibinfo {volume} {101}},\ \bibinfo
  {pages} {104201} (\bibinfo {year} {2020})}\BibitemShut {NoStop}%
\bibitem [{Note1()}]{Note1}%
  \BibitemOpen
  \bibinfo {note} {It should be noted that using the joint distribution for
  uncorrelated levels i.e. $P(E_n,n;E_m,m) = P(E_m) P(E_n)$ would produce a
  different SFF. See the supplementary materials~\cite {supp} for more details
  about this and the relationship to the SFF form in the main
  text.}\BibitemShut {Stop}%
\bibitem [{\citenamefont {Jia}\ and\ \citenamefont
  {Verbaarschot}(2020)}]{JiaVerbaaschot_2020}%
  \BibitemOpen
  \bibfield  {author} {\bibinfo {author} {\bibfnamefont {Y.}~\bibnamefont
  {Jia}}\ and\ \bibinfo {author} {\bibfnamefont {J.~J.~M.}\ \bibnamefont
  {Verbaarschot}},\ }\bibfield  {title} {\bibinfo {title} {Spectral
  fluctuations in the sachdev-ye-kitaev model},\ }\bibfield  {journal}
  {\bibinfo  {journal} {Journal of High Energy Physics}\ }\textbf {\bibinfo
  {volume} {2020}},\ \href {https://doi.org/10.1007/jhep07(2020)193}
  {10.1007/jhep07(2020)193} (\bibinfo {year} {2020})\BibitemShut {NoStop}%
\bibitem [{\citenamefont {Riser}\ and\ \citenamefont
  {Kanzieper}(2021)}]{riser2020power}%
  \BibitemOpen
  \bibfield  {author} {\bibinfo {author} {\bibfnamefont {R.}~\bibnamefont
  {Riser}}\ and\ \bibinfo {author} {\bibfnamefont {E.}~\bibnamefont
  {Kanzieper}},\ }\bibfield  {title} {\bibinfo {title} {Power spectrum and form
  factor in random diagonal matrices and integrable billiards},\ }\href
  {https://doi.org/https://doi.org/10.1016/j.aop.2020.168393} {\bibfield
  {journal} {\bibinfo  {journal} {Annals of Physics}\ }\textbf {\bibinfo
  {volume} {425}},\ \bibinfo {pages} {168393} (\bibinfo {year}
  {2021})}\BibitemShut {NoStop}%
\end{thebibliography}%


\begin{thebibliography}{5}%
\makeatletter
\providecommand \@ifxundefined [1]{%
 \@ifx{#1\undefined}
}%
\providecommand \@ifnum [1]{%
 \ifnum #1\expandafter \@firstoftwo
 \else \expandafter \@secondoftwo
 \fi
}%
\providecommand \@ifx [1]{%
 \ifx #1\expandafter \@firstoftwo
 \else \expandafter \@secondoftwo
 \fi
}%
\providecommand \natexlab [1]{#1}%
\providecommand \enquote  [1]{``#1''}%
\providecommand \bibnamefont  [1]{#1}%
\providecommand \bibfnamefont [1]{#1}%
\providecommand \citenamefont [1]{#1}%
\providecommand \href@noop [0]{\@secondoftwo}%
\providecommand \href [0]{\begingroup \@sanitize@url \@href}%
\providecommand \@href[1]{\@@startlink{#1}\@@href}%
\providecommand \@@href[1]{\endgroup#1\@@endlink}%
\providecommand \@sanitize@url [0]{\catcode `\\12\catcode `\$12\catcode
  `\&12\catcode `\#12\catcode `\^12\catcode `\_12\catcode `\%12\relax}%
\providecommand \@@startlink[1]{}%
\providecommand \@@endlink[0]{}%
\providecommand \url  [0]{\begingroup\@sanitize@url \@url }%
\providecommand \@url [1]{\endgroup\@href {#1}{\urlprefix }}%
\providecommand \urlprefix  [0]{URL }%
\providecommand \Eprint [0]{\href }%
\providecommand \doibase [0]{https://doi.org/}%
\providecommand \selectlanguage [0]{\@gobble}%
\providecommand \bibinfo  [0]{\@secondoftwo}%
\providecommand \bibfield  [0]{\@secondoftwo}%
\providecommand \translation [1]{[#1]}%
\providecommand \BibitemOpen [0]{}%
\providecommand \bibitemStop [0]{}%
\providecommand \bibitemNoStop [0]{.\EOS\space}%
\providecommand \EOS [0]{\spacefactor3000\relax}%
\providecommand \BibitemShut  [1]{\csname bibitem#1\endcsname}%
\let\auto@bib@innerbib\@empty
\bibitem [{\citenamefont {Riser}\ \emph {et~al.}(2020)\citenamefont {Riser},
  \citenamefont {Osipov},\ and\ \citenamefont
  {Kanzieper}}]{riser2020nonperturbative}%
  \BibitemOpen
  \bibfield  {author} {\bibinfo {author} {\bibfnamefont {R.}~\bibnamefont
  {Riser}}, \bibinfo {author} {\bibfnamefont {V.~A.}\ \bibnamefont {Osipov}},\
  and\ \bibinfo {author} {\bibfnamefont {E.}~\bibnamefont {Kanzieper}},\
  }\bibfield  {title} {\bibinfo {title} {Nonperturbative theory of power
  spectrum in complex systems},\ }\href@noop {} {\bibfield  {journal} {\bibinfo
   {journal} {Annals of Physics}\ }\textbf {\bibinfo {volume} {413}},\ \bibinfo
  {pages} {168065} (\bibinfo {year} {2020})}\BibitemShut {NoStop}%
\bibitem [{\citenamefont {Berry}\ and\ \citenamefont
  {Tabor}(1977)}]{BerryTabor1977_level}%
  \BibitemOpen
  \bibfield  {author} {\bibinfo {author} {\bibfnamefont {M.~V.}\ \bibnamefont
  {Berry}}\ and\ \bibinfo {author} {\bibfnamefont {M.}~\bibnamefont {Tabor}},\
  }\bibfield  {title} {\bibinfo {title} {Level clustering in the regular
  spectrum},\ }\href@noop {} {\bibfield  {journal} {\bibinfo  {journal}
  {Proceedings of the Royal Society of London. A. Mathematical and Physical
  Sciences}\ }\textbf {\bibinfo {volume} {356}},\ \bibinfo {pages} {375}
  (\bibinfo {year} {1977})}\BibitemShut {NoStop}%
\bibitem [{\citenamefont {Marklof}(2001)}]{BerryTabor_conjecture_Jens}%
  \BibitemOpen
  \bibfield  {author} {\bibinfo {author} {\bibfnamefont {J.}~\bibnamefont
  {Marklof}},\ }\bibfield  {title} {\bibinfo {title} {The berry-tabor
  conjecture},\ }in\ \href@noop {} {\emph {\bibinfo {booktitle} {European
  Congress of Mathematics}}},\ \bibinfo {editor} {edited by\ \bibinfo {editor}
  {\bibfnamefont {C.}~\bibnamefont {Casacuberta}}, \bibinfo {editor}
  {\bibfnamefont {R.~M.}\ \bibnamefont {Mir{\'o}-Roig}}, \bibinfo {editor}
  {\bibfnamefont {J.}~\bibnamefont {Verdera}},\ and\ \bibinfo {editor}
  {\bibfnamefont {S.}~\bibnamefont {Xamb{\'o}-Descamps}}}\ (\bibinfo
  {publisher} {Birkh{\"a}user Basel},\ \bibinfo {address} {Basel},\ \bibinfo
  {year} {2001})\ pp.\ \bibinfo {pages} {421--427}\BibitemShut {NoStop}%
\bibitem [{\citenamefont {Oganesyan}\ and\ \citenamefont
  {Huse}(2007)}]{OganesyanHuse_2007_PhysRevB.75.155111}%
  \BibitemOpen
  \bibfield  {author} {\bibinfo {author} {\bibfnamefont {V.}~\bibnamefont
  {Oganesyan}}\ and\ \bibinfo {author} {\bibfnamefont {D.~A.}\ \bibnamefont
  {Huse}},\ }\bibfield  {title} {\bibinfo {title} {Localization of interacting
  fermions at high temperature},\ }\href
  {https://doi.org/10.1103/PhysRevB.75.155111} {\bibfield  {journal} {\bibinfo
  {journal} {Phys. Rev. B}\ }\textbf {\bibinfo {volume} {75}},\ \bibinfo
  {pages} {155111} (\bibinfo {year} {2007})}\BibitemShut {NoStop}%
\bibitem [{\citenamefont {Balakrishnan}(2003)}]{balakrishnan2003all}%
  \BibitemOpen
  \bibfield  {author} {\bibinfo {author} {\bibfnamefont {V.}~\bibnamefont
  {Balakrishnan}},\ }\bibfield  {title} {\bibinfo {title} {All about the dirac
  delta function},\ }\href@noop {} {\bibfield  {journal} {\bibinfo  {journal}
  {Resonance}\ }\textbf {\bibinfo {volume} {8}},\ \bibinfo {pages} {48}
  (\bibinfo {year} {2003})}\BibitemShut {NoStop}%
\end{thebibliography}%

\end{document}


\title{Supplementary Material: The universal spectral form factor for many-body localization}

\author{Abhishodh Prakash}
\email{abhishodh.prakash@icts.res.in}
\affiliation{International Centre for Theoretical Sciences (ICTS-TIFR),
Tata Institute of Fundamental Research,
Shivakote, Hesaraghatta Hobli,
Bengaluru 560089, India}
\author{J. H. Pixley}
\email{jed.pixley@physics.rutgers.edu}
\affiliation{Department of Physics and Astronomy, Center for Materials Theory, Rutgers University, Piscataway, NJ 08854 USA}
\author{Manas Kulkarni}
\email{manas.kulkarni@icts.res.in}
\affiliation{International Centre for Theoretical Sciences (ICTS-TIFR),
Tata Institute of Fundamental Research,
Shivakote, Hesaraghatta Hobli,
Bengaluru 560089, India}

\date{\today}

\maketitle

\tableofcontents

\section{S1. SFF for Poisson numbers}
We now provide details of the derivation of the expressions for the spectral form factors (SFF) of Poisson levels presented in the main text. Consider levels from a Poisson process $\{E_i\}$ generated by summing gaps $\{\delta_i\}$ with an exponential i.i.d  
\begin{eqnarray}
P(\{\delta_i\}) &=& \prod_i \rho(\delta_i),\\
\rho(\delta) &=& \frac{1}{\mu} \exp \left(-\frac{\delta}{\mu}\right),\\
E_i &=& \sum_{k=1}^i \delta_k.
\end{eqnarray}
To compute the SFF,  we need the joint two-point distribution $P(E_n,n;E_m,m)$ i.e. the probability that the $m^{th}$ eigenvalue is $E_m$ and the $n^{th}$ eigenvalue is $E_n$. Assuming with no loss of generality $m>n$, this can be obtained as follows
\begin{align}
P(E_n,n;E_m,m) = p(E_n,n)~ p(E_m-E_n,m-n),
\end{align}
where $p(E_k,k)$ is the well known Poisson distribution
\begin{equation}
p(E_k,k) =   \frac{e^{-\frac{E_k}{\mu} }}{\mu (k-1)!} \left(\frac{E_k}{\mu}\right)^{k-1}.
\end{equation}
We assume $E_k>0$ and $k = 1,2,3,\ldots$. Let us sketch the derivation of the above equation
\begin{align}
P(E_n,n;E_m,m) &=    \int_{0}^{\infty} d \delta_1 \rho(\delta_1) \cdots \int_{0}^{\infty} d \delta_N \rho(\delta_N)~  \delta\left(E_m-\sum_{i=1}^m \delta_i \right) ~\delta\left(E_n-\sum_{i=1}^n \delta_i \right) \nonumber\\
&=    \int_{0}^{\infty} d \delta_1 \rho(\delta_1) \cdots \int_{0}^{\infty} d \delta_N \rho(\delta_N)~ \int_{-\infty}^{\infty} \frac{dk}{2 \pi}~ e^{ik\left(E_m-\sum_{i=1}^m \delta_i \right)} ~ \int_{-\infty}^{\infty} \frac{dk'}{2 \pi}~ e^{ik'\left(E_n-\sum_{i=1}^n \delta_i \right)}\nonumber \\
&= \int_{-\infty}^{\infty} \frac{dk}{2 \pi}  \int_{-\infty}^{\infty} \frac{dk'}{2 \pi} ~e^{i(k E_m+k' E_n)} \left[\prod_{l=1}^n \int_{0}^\infty d \delta_l~ \rho(\delta_l)~ e^{-i (k+k') \delta_l}\right] \left[\prod_{l={n+1}}^m \int_{0}^\infty d \delta_l~ \rho(\delta_l)~ e^{-i k \delta_l}\right] \nonumber\\
&= \int_{-\infty}^{\infty} \frac{dk}{2 \pi}  \int_{-\infty}^{\infty} \frac{dk'}{2 \pi} ~e^{i(k E_m+k' E_n)} \left[\moy{e^{-i (k+k') \delta}}_\rho\right]^n ~\left[\moy{e^{-i k \delta}}_\rho\right]^{m-n}.
\end{align}
We now change variables $\{k,k'\} \mapsto \{k,l=k+k'\}$
\begin{align}
P(E_n,n;E_m,m) &= \int_{-\infty}^{\infty} \frac{dk}{2 \pi} ~e^{ik (E_m-E_n)}  \left[\moy{e^{-i k \delta}}_\rho\right]^{m-n} \int_{-\infty}^{\infty} \frac{dl}{2 \pi} ~e^{il E_n} \left[\moy{e^{-i l \delta}}_\rho\right]^n \nonumber \\
&= p(E_n,n) ~p(E_m-E_n,m-n),
\end{align}
where
\begin{align}
\moy{e^{-i k \delta}}_\rho &=  \int_{0}^\infty d \delta~ \rho(\delta)~ e^{-i k \delta} = \frac{1}{1+i\mu k},\\
p(E_n,n) &= \int_{-\infty}^{\infty} \frac{dk}{2 \pi} ~e^{ik E_n} \left[\moy{e^{-i k \delta}}_\rho\right]^n, \nonumber\\ &= \int_{-\infty}^{\infty} \frac{dk}{2 \pi} ~ \frac{e^{ik E_n}}{(1+i\mu k)^n} =  \frac{e^{-\frac{E_n}{\mu} }}{\mu (n-1)!} \left(\frac{E_n}{\mu}\right)^{n-1}.  
\end{align}
In the last step, the integral is performed by closing the contour in the upper-half complex plane to enclose the $n^{th}$ order pole at $k=i/\mu $ and using Cauchy's integral formula.

\begin{figure}[!h]
	\centering
	\begin{tabular}{cc}
		\includegraphics[width=0.4\textwidth]{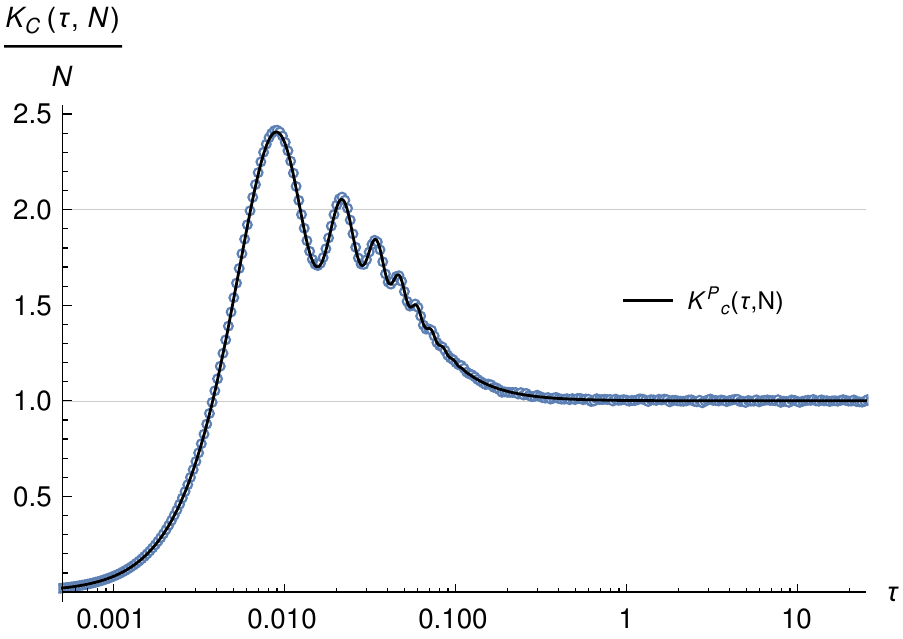} &
		\includegraphics[width=0.4\textwidth]{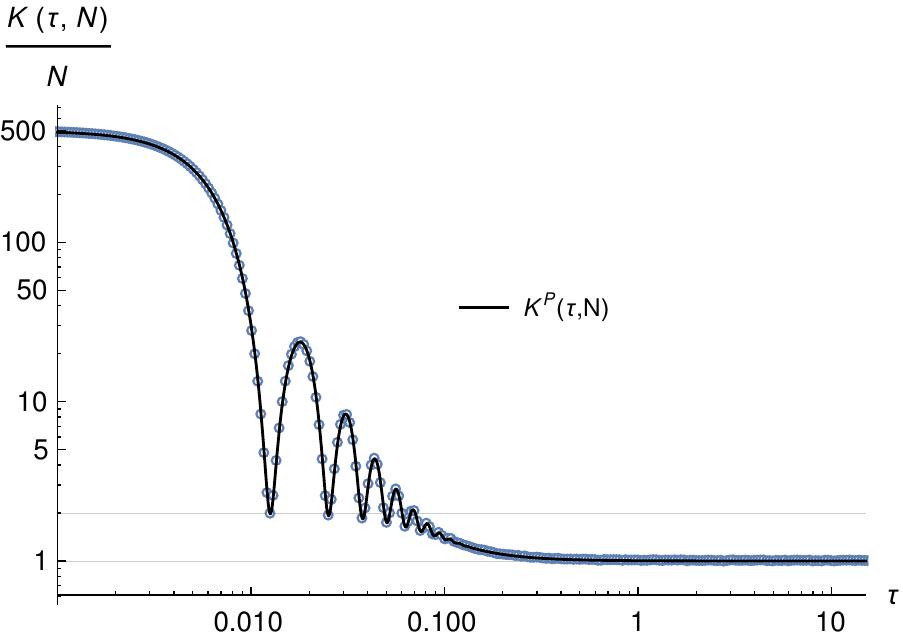} 
	\end{tabular}
	\caption{The connected SFF (left) and SFF (right) for $N=500$ numerically generated Poisson levels with unit mean level spacing averaged over 50,000 realizations. The analytic result is shown as the black line.}
	\label{fig:SFF_Poissoncheck}
\end{figure}

Let us now proceed to compute the SFF
\begin{align}
K(\tau,N) &= \sum_{m,n =1}^N \langle e^{i \tau (E_m - E_n)} \rangle_P 
= \sum_{m=n=1}^N 1 + \sum_{m \neq n =1}^N \langle e^{i \tau (E_m - E_n)} \rangle_P
\nonumber \\ &= N + \sum_{n=1}^N \sum_{m = n+1}^N \left(\langle e^{i \tau (E_m - E_n)} \rangle_P + \langle e^{-i \tau (E_m - E_n)} \rangle_P\right),  \label{eq:SFF_intermediate_Poisson}
\end{align}
where
\begin{align}
\langle e^{i \tau (E_m - E_n)}  \rangle_P &= \int_{0}^\infty dE_m~\int_{0}^\infty dE_n~ P(E_n,n;E_m,m) ~e^{i \tau (E_m - E_n)} \nonumber \\
&= \int_{0}^\infty dE_n~ p(E_n,n) \int_{0}^\infty  dE_m~ p(E_m-E_n,m-n) ~e^{i \tau (E_m - E_n)}.
\end{align}
Changing the integration variables $\{E_m, E_n\} \mapsto \{E_{mn} = E_m-E_n, E_n\}$, we get
\begin{align}
\langle e^{i \tau (E_m - E_n)}  \rangle_P &= \int_{0}^\infty dE_n~ p(E_n,n) \int_{0}^\infty  dE_{mn}~ p(E_{mn},m-n) ~e^{i \tau E_{mn}} \nonumber \\
& = \int_{0}^\infty  dE_{mn}~ p(E_{mn},m-n) ~e^{i \tau E_{mn}}  = \frac{1}{(1-i \mu \tau)^{m-n}},
\end{align}
where we have used $\int_{0}^\infty dE_n~ p(E_n,n) =1$. Substituting the above expression in \cref{eq:SFF_intermediate_Poisson} gives
\begin{align}
K(\tau,N) &=  N + \sum_{n=1}^N \sum_{m = n+1}^N \left[\frac{1}{(1+i \mu \tau)^{m-n}} + \frac{1}{(1-i \mu \tau)^{m-n}}\right] .
\end{align}
The summand of the above series depends only on the differences $k=m-n$. As a result, we can change the summation over $m$ and $n$ in terms of $k$ as follows
\begin{align}
K(\tau,N) = N + \sum_{k=1}^{N-1} (N-k) \left[\frac{1}{(1+i \mu \tau)^k} + \frac{1}{(1-i \mu \tau)^k}\right].
\end{align}
The above series can be summed up easily using standard results from summing the geometric series 
\begin{align}
S^G_n = \sum_{k=1}^n br^{k-1} = b + br + br^2 + \ldots + br^{n-1} =  \frac{b \left(1-r^n\right)}{\left(1-r\right),}
\end{align}
and the arithmetico-geometric series 
\begin{align}
S_n &= \sum_{k=1}^n \left(a + (k-1)d\right) br^{k-1} =  ab + (a+d) br + (a+2d) br^2 + \ldots + (a+ (n-1)d) br^{n-1} \nonumber \\
&= \frac{ab}{1-r} + \frac{dbr (1-r^n)}{(1-r)^2} - \frac{(a+nd) br^n}{(1-r)} ,
\end{align}
with $a=d=1$, $b=r=\frac{1}{(1 \pm i \mu \tau)}$ and $n=N-1$ to get the final answer
\begin{equation}
K(\tau,N) = N + \frac{2}{(\mu\tau)^2}  - \frac{ (1+i \mu \tau)^{1-N} + (1-i \mu \tau)^{1-N}  }{(\mu\tau)^2}. \label{eq:SFF_Poisson2}
\end{equation}
We now proceed to calculate the connected SFF 
\begin{equation}
K_c(\tau,N) = K(\tau,N) -  |\sum_{m} \langle e^{i \tau E_m}  \rangle_P|^2,
\label{eq:connected_intermediate}
\end{equation}
where
\begin{align}
\langle e^{i \tau E_m} \rangle_P &=  \int_{0}^\infty  dE_m~ p(E_m,m) ~e^{i \tau E_m} = \frac{1}{(1-i \mu \tau)^{m}}, \\
\sum_{m=1}^N \langle e^{i \tau E_m} \rangle_P &= 
\sum_{m=1}^N \frac{1}{(1-i \mu \tau )^m}  = \frac{(1-i \mu\tau)^{-N}-1}{i \mu \tau}.
\end{align}
Plugging this into \cref{eq:connected_intermediate} and some simplification, we get the final answer
\begin{equation}
K_c(\tau,N) 
= N + \frac{1}{(\mu\tau)^2} - \frac{(1+(\mu\tau)^2)^{-N}}{(\mu\tau)^2}  - \frac{i}{  \mu \tau } \left[ (1 + i \mu \tau )^{-N} - (1 - i \mu \tau )^{-N}  \right] . \label{eq:SFFC_Poisson2}
\end{equation}
\Cref{fig:SFF_Poissoncheck} shows the SFF computed numerically for $N$ Poisson levels and the excellent  agreement with the above analytical result. Note that the expression for the SFF was obtained by the authors of \cite{riser2020nonperturbative} as a special case of a more general result for uncorrelated gap distributions.

\section{S2. Various time-scales for the SFF}
Let us now closely analyze the various time-scales of the SFF for Poisson levels. Let us focus on the connected SFF for concreteness and clarity. We also assume that $N \gg 1$ although finite. Let us start with the expression for $K^P_c(\tau,N)$
\begin{equation}
K^P_c(\tau,N) 
= N + \frac{1}{(\mu\tau)^2} - \frac{(1+(\mu\tau)^2)^{-N}}{(\mu\tau)^2}  - \frac{i}{  \mu \tau } \left[ (1 + i \mu \tau )^{-N} - (1 - i \mu \tau )^{-N}  \right] 
\end{equation}
For large values of $ \mu \tau \gg 1$, we have
\begin{align}
K^{late}_c(\tau,N) \approx N.
\end{align}
For $ \frac{1}{\sqrt{N}} <  \mu \tau <1$, we have the result stated in the main-text
\begin{align}
K^{inter}_c(\tau,N) \approx N + \frac{1}{(\mu \tau)^2}.
\label{eq:SFF_intermediate_univrsal}
\end{align}  
For $ \mu\tau < \frac{1} {\sqrt{N}}$, we can get the form of the SFF by a careful Taylor expansion as follows
\begin{align}
K^{early}_c(\tau,N) &= N + \frac{1}{(\mu\tau)^2} - \frac{(1+(\mu\tau)^2)^{-N}}{(\mu\tau)^2} -  2 \left(1+(\mu\tau)^2\right)^{-
\frac{N}{2}}\frac{ \sin \left(N \tan^{-1}(\mu \tau)\right)}{(\mu \tau)} \nonumber\\
&\approx N + \frac{1}{(\mu\tau)^2} - \frac{(1-N(\mu\tau)^2)}{(\mu\tau)^2} -  2\left(1-\frac{N}{2}(\mu\tau)^2\right) \frac{ \sin \left(N \mu \tau\right)}{(\mu \tau)} \nonumber \\
&\approx 2N \left(1 -\frac{ \sin \left(N \mu \tau\right)}{(N\mu \tau)} \right).
\end{align}
In this small $\tau$ regime, oscillations with a period $T=2\pi/(N\mu)$ appear.
\begin{figure}[!h]
	\centering
	\begin{tabular}{cc}
		\includegraphics[width=0.45\textwidth]{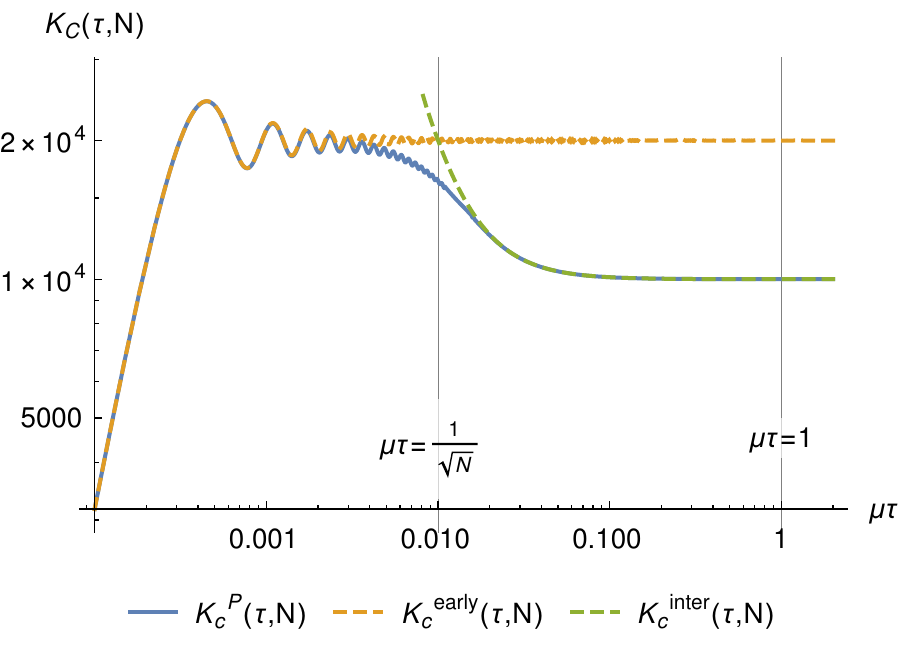} & 
			\includegraphics[width=0.45\textwidth]{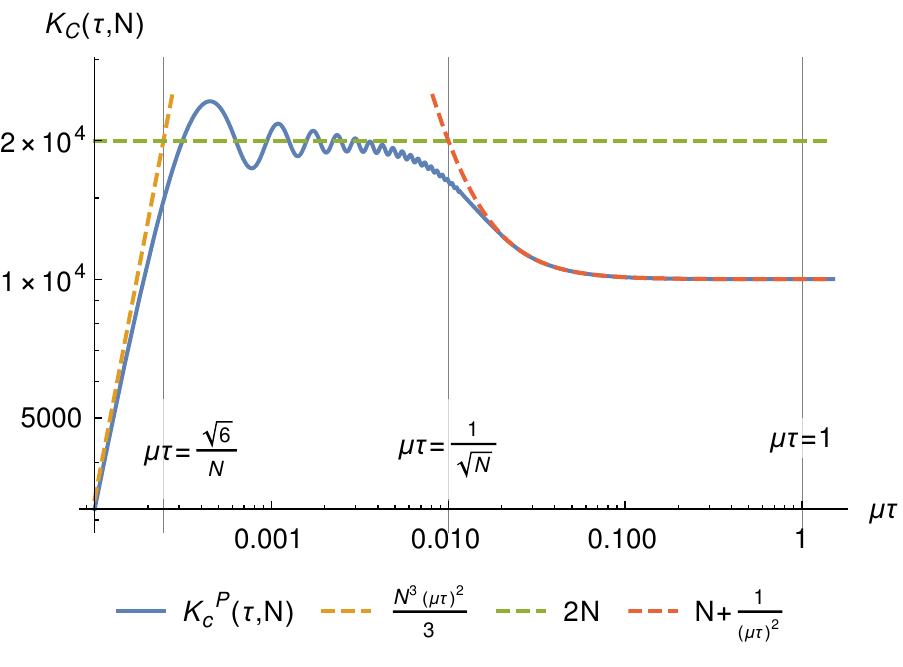} 
	\end{tabular}
	\caption{ The connected SFF with (left) the early and intermediate forms separated and (right) the various $\tau$ regimes separated \label{fig:timescales}.}
\end{figure}

The early-$\tau$ SFF can be further separated into two regions. For $ \frac{1}{N} < \mu \tau < \frac{1}{\sqrt{N}}$, we have $K_c^{early} \approx 2N$. For  $\mu \tau < \frac{1}{N}$, the sine term can be Taylor expanded to give us $K^{early}_c \approx \frac{N^3 (\mu \tau)^2}{3}$. The quadratic and constant pieces meet at $\mu \tau = \frac{\sqrt{6}}{N}$. Putting all these together, we see that for finite but large N, the connected SFF can be divided into four $\tau$ regimes as shown in~\cref{fig:timescales}. 

Note that similar time-scales as well as scaling forms were also obtained in \cite{riser2020nonperturbative} although their notion of universality (independence of underlying gap distribution) is  different from ours. While we consider the early $\tau$ regime to be non-universal since it is affected by the global density of states, \cite{riser2020nonperturbative} considers the same to be universal since the expressions assume a form that is independent of the underlying distribution of the uncorrelated gaps. 

\section{S3. SFF for completely uncorrelated numbers to Poisson numbers}
The often-quoted relationship between Poisson processes and uncorrelated numbers is clarified in this section. First, we derive the SFF for uncorrelated numbers and demonstrate that it is different from the form obtained previously in~\cref{eq:SFF_Poisson2,eq:SFFC_Poisson2}. Next, we present the setting when the SFF for uncorrelated numbers reduces to that of the Poisson process.

\subsection{S3a. SFF for completely uncorrelated levels (random diagonal ensemble)}
Consider an $N \times N$ matrix with random entries along the diagonal drawn from some independent and identical distribution (i.i.d) $p(E)$. The eigenvalues $\{E_i\}$ of such a matrix is simply the diagonal entries sorted. As can be easily seen from the formulas for the SFF,
\begin{align}
K(\tau,N) &=  \langle \sum_{m,n=1}^N e^{i \tau (E_m - E_n)} \rangle, \\
K_c(\tau,N) &=  \langle \sum_{m,n=1}^N e^{i \tau (E_m - E_n)}   \rangle - |\sum_{m=1}^N \langle e^{i \tau E_m}   \rangle|^2,
\end{align}
the ordering of $\{E_i\}$ is irrelevant to compute the SFF. We need the joint two-point distribution $P(E_n,n;E_m,m)$, i.e. probability that the $m^{th}$ eigenvalue is $E_m$ and the $n^{th}$ eigenvalue is $E_n$. Since we can ignore ordering, this is simply
\begin{equation}
P(E_n,n;E_m,m) =  p(E_n) ~p(E_m).
\label{eq:P2 random diagonal}
\end{equation}
Let us start with the disconnected SFF for random levels, $K^R(\tau,N)$
\begin{align}
K^R(\tau,N) &= \sum_{m,n}  \langle e^{i \tau (E_m - E_n)} \rangle 
= \sum_{m=n} 1 + \sum_{m \neq n} \langle e^{i \tau E_m } e^{-i \tau  E_n}  \rangle_P  = N + N(N-1) |\langle e^{i \tau E}  \rangle_p|^2,
\end{align}
where we have used 
\begin{eqnarray}
\langle e^{i \tau E_m } e^{-i \tau  E_n} \rangle_P &=& \langle e^{i \tau E_m }  \rangle_p \langle e^{-i \tau  E_n} \rangle_p, \\
\text{and }\langle e^{i \tau E_{m}}\rangle
&=&
\langle e^{i \tau E_{n}}
\rangle_p = \langle e^{i \tau E} \rangle_p = \int_0^{2D} dE ~p(E) ~e^{i \tau E}.
\end{eqnarray}
We can also get the connected SFF easily
\begin{equation}
K^R_c(\tau,N) = K(\tau,N) -  |\sum_{m} \langle e^{i \tau E_m}  \rangle|^2 
= K(\tau,N) - N^2 |\langle e^{i \tau E}  \rangle|^2 
= N(1- |\langle e^{i \tau E}  \rangle|^2), \label{eq:Sffc random level}
\end{equation}
\begin{figure}[!h]
	\centering
	\begin{tabular}{cc}
		\includegraphics[width=0.4\textwidth]{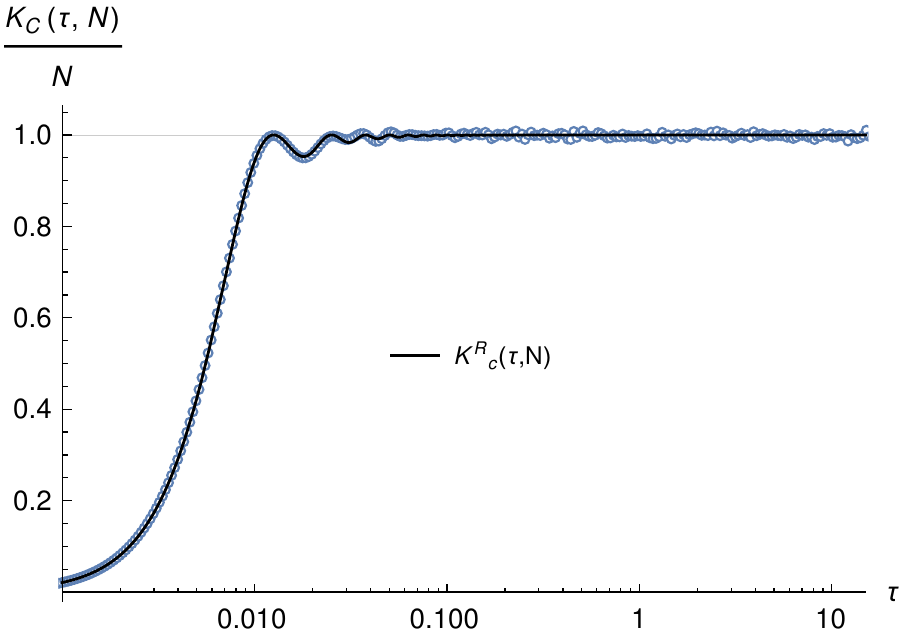} &
		\includegraphics[width=0.4\textwidth]{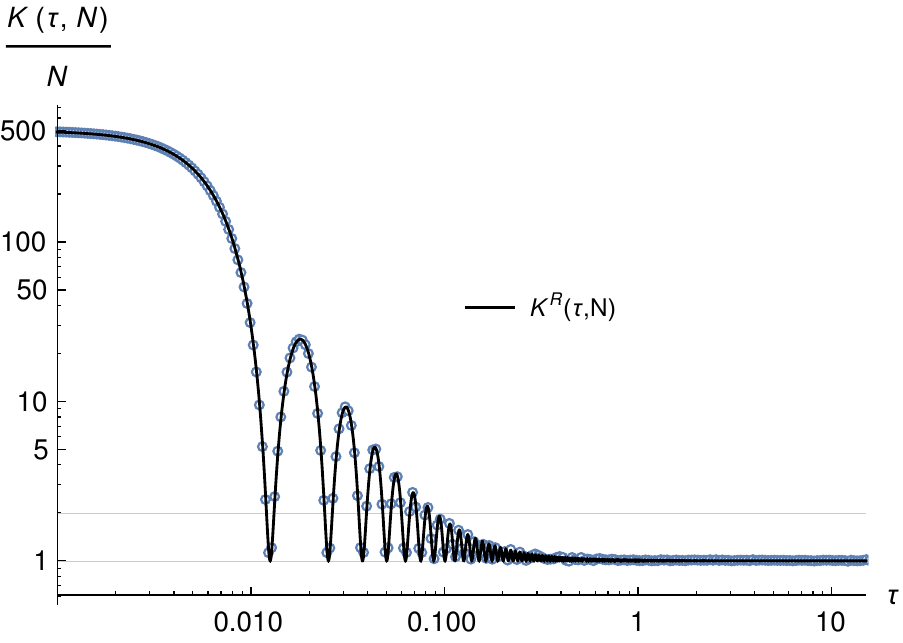} 
	\end{tabular}
	\caption{The connected (left) and disconnected (right) SFF for random levels with unit mean level spacing. The analytic result is the black line for $N=500$ levels, which we compare with the numerically computed SFF using levels drawn from a uniform distribution that is averaged over 50,000  realizations.}
	\label{fig:SFF_Randomcheck}
\end{figure}
As a concrete example, if the levels are drawn from a uniform distribution of width $D$ i.e.
\begin{align}
p(E) &= \begin{cases}
\frac{1}{D} ~for~ 0 \le E \le D\\
0 ~~otherwise
\end{cases},
\end{align}
the expressions for SFF are as follows
\begin{align}
K^R(\tau,N) &= N + N(N-1) \left| \frac{\sin(D \tau/2)}{(D \tau/2)} \right|^2, \\
K^R_c(\tau,N) & = N\left(1- \left| \frac{\sin(D \tau/2)}{(D \tau/2)} \right|^2\right).
\end{align}
It is more convenient to express these results in terms of the number of levels $N$ and the mean level spacing $\mu$ which are related as $D = \mu N$.
\begin{align}
K^R(\tau,N) &= N + N(N-1) \left| \frac{\sin(\mu N \tau/2)}{(\mu N \tau/2)} \right|^2, \label{eq:Uniform SFF} \\
K^R_c(\tau,N) & = N\left(1- \left| \frac{\sin(\mu N \tau/2)}{(\mu N \tau/2)} \right|^2\right)\label{eq:Uniform SFFc}.
\end{align}
\Cref{fig:SFF_Randomcheck} shows a comparison of the analytic  result for the SFF for random levels and with it, the numerically computed SFF for $N$ random numbers drawn from a uniform distribution with width $D=N$ (to ensure unit mean level spacing) and its excellent matching with the above analytical expression. Observe that $K_c(\tau,N)$ in \cref{eq:Sffc random level} strictly satisfies the inequality.
\begin{align}
    K^R_c(\tau,N) \le N
    \label{eq:random_inequality}
\end{align}
Recall that for Poisson levels, for large $N$ and intermediate values of $\tau$ we have 
\begin{align}
    K^P_c(\tau,N) \approx N + \frac{1}{(\mu \tau)^2} \ge N.
\end{align}
Thus, the sub leading universal $\frac{1}{(\mu \tau)^2}$ form, which violates the inequality, is impossible to be present in $K^R_c(\tau,N)$. We conclude that the SFF for completely uncorrelated random levels is different from the SFF for Poisson numbers, $K^P_c(\tau,N)$!

\subsection{S3b. From Random levels to Poisson process}
As we have seen, the SFF for numbers drawn from a Poisson process is different from that of uncorrelated levels. In this section, we review the relationship between uncorrelated levels and Poisson processes. We specify the condition under which the SFF of the former reduces to the latter. 

\begin{figure}[!h]
	\centering
	\begin{tabular}{c}
		\includegraphics[width=0.5\textwidth]{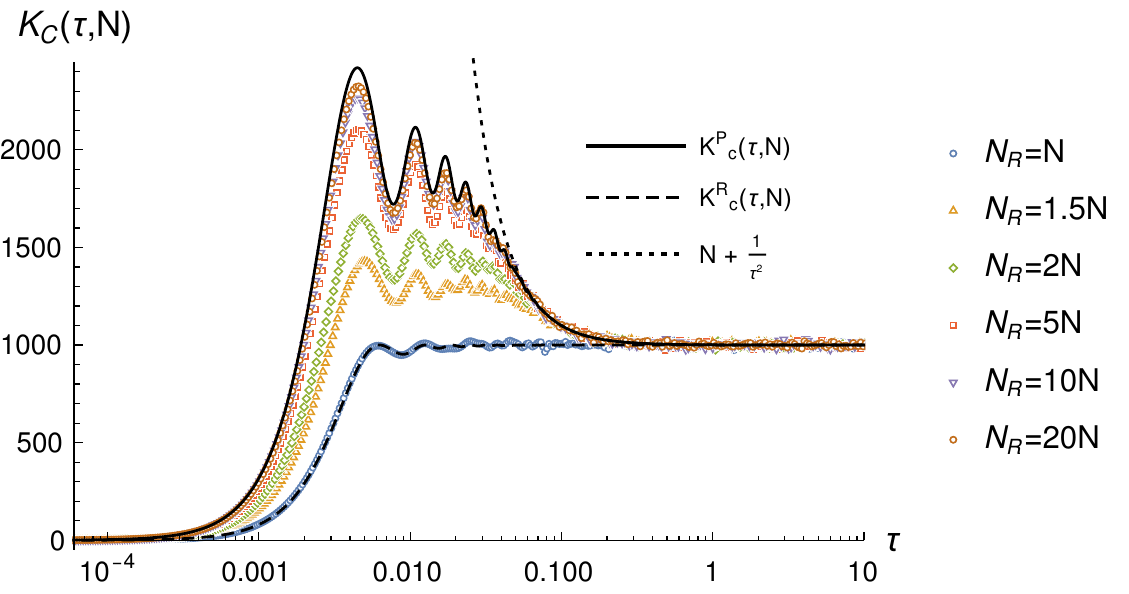} 
	\end{tabular}
	\caption{The evolution of the connected SFF from Random~\cref{eq:Uniform SFFc} to Poisson~\cref{eq:SFFC_Poisson2}. $N=1000$ levels chosen from an ensemble of sorted $N_R$ numbers with drawn from a uniform distribution with unit mean level spacing and averaged over $10^5$ disorder realizations. \label{fig:PoissontoRandom}.}
\end{figure}

Let us once again consider $N$ uncorrelated random levels drawn from any distribution. Let $E_1, E_2, \ldots, E_N$ be the same random numbers after sorting. The process of sorting introduces a weak correlation that can be observed in local spectral statistics. For example, the mean level spacing distribution,
\begin{equation}
\moy{P(s,N)} =  \frac{1}{N} \sum_i \moy{\delta(s-E_{i+1}+ E_i)},
\end{equation}
approaches the Poisson one 
\begin{equation}
\moy{P(s,N)}  \xrightarrow{N \gg 1 } P(s) = \frac{1}{\mu} e^{-\frac{s}{\mu}},
\end{equation}
where $\mu$ is inherited from the original distribution of the random numbers. This also extends to the distribution of larger spacings i.e. 
\begin{equation}
\moy{P(k,s,N)} =  \frac{1}{N} \sum_i \moy{\delta(s-E_{i+k}+ E_i)} \xrightarrow{N\gg k } P(k,s) =  \frac{s^{k-1}}{\mu^k (k-1)!} e^{-\frac{s}{\mu}}.
\end{equation}
\begin{figure}[!htbp]
	\centering
	\begin{tabular}{cc}
		\includegraphics[width=0.4\textwidth]{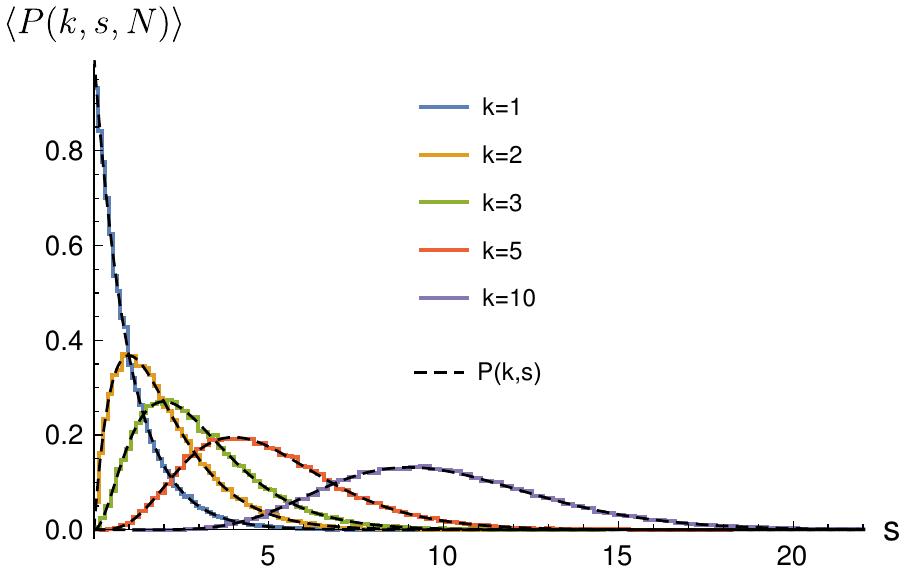} & 		
		\includegraphics[width=0.4\textwidth]{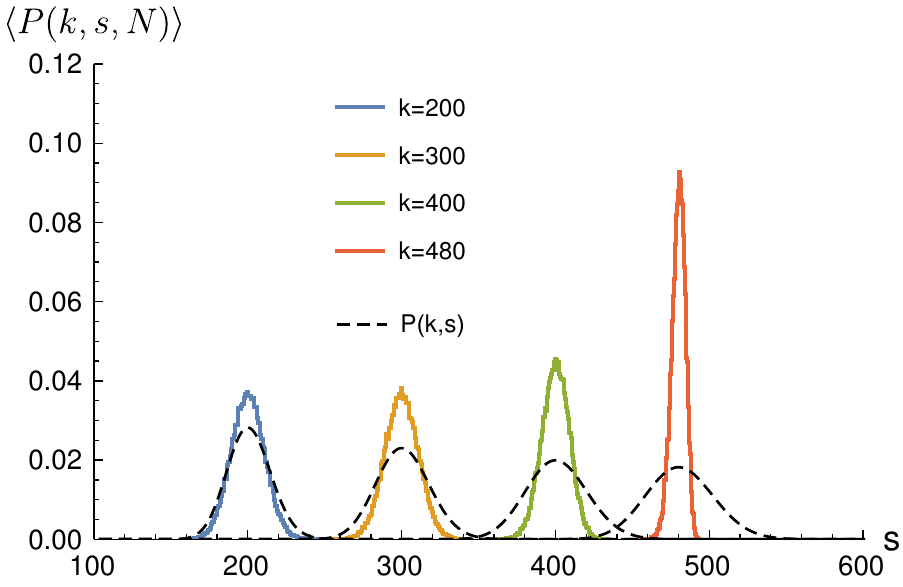} 
	\end{tabular}	
	\caption{$\moy{P(k,s,N)}$ for various values of $k$ numerically computed for $N=500$ sorted random numbers drawn from a uniform distribution with mean level spacing $\mu=1$ averaged over $50,000$ samples. It can be seen that $\moy{P(k,s,N)}$ agrees with the Poisson distribution, $P(k,s)$ (dashed lines) for small values of $k$ but deviates increasingly when $k \sim N$. \label{fig:Poisson from uncorrelated}}
\end{figure}
Note that crucially, the above is valid only for $k\ll N$. For $k \sim N$ on the other hand, the distribution of $\moy{P(k,s,N)}$ would depend on the underlying i.i.d distribution of the uncorrelated numbers. These facts can be easily checked numerically as shown in~\cref{fig:Poisson from uncorrelated}. In the limit $N \rightarrow \infty$, we can expect the Poisson distribution for any finite $k$. We can now understand why the SFF for random levels is different from that of numbers from a Poisson process. Given uncorrelated random levels, the SFF is computed using \emph{all} differences in eigenvalues $E_i - E_j$. When $|i-j| \ll N$, the levels are Poissonian but when $|i-j| \sim N$, they are not. As a result, the total SFF, that sums over \emph{all} differences $E_i - E_j$ deviates for Random levels from the Poisson form. However, we can recover the Poisson SFF in the following way. Consider an ensemble of $N_R$ uncorrelated random numbers and sort them (eigenvalues of $N_R \times N_R$ random diagonal ensemble). From each sample, let us then choose only $N$ levels (from the middle, say) out of $N_R$ levels to compute the SFF. If $N \ll N_R$, the levels resemble a Poisson process in the sense all differences $E_i - E_j$ follow a Poisson distribution and we should recover the Poisson SFF. As $N_R \rightarrow \infty$, we should get the Poisson SFF for any $N$. This can indeed be checked to be true numerically and shown in \cref{fig:PoissontoRandom} for the connected SFF where the evolution from random to Poisson SFF can be clearly seen. Starting from $N=N_R$ where the SFF matches with the form~\cref{eq:Uniform SFFc}, with increasing $N_R$, the SFF form moves closer to the Poisson one.

These results can be summarized by taking two different limits of the spectral form factor, namely for $N_R\gg 1$ we have
\begin{equation}
\lim_{N\rightarrow N_R} K(\tau,N) \rightarrow K^R(\tau,N) =  N+N(N-1) \left| \frac{\sin(D \tau/2)}{(D \tau/2)} \right|^2
\end{equation}
where $N=N_R=D$ and
\begin{equation}
    \lim_{N/N_R\rightarrow 0}K(\tau,N)\rightarrow K^P(\tau,N)= N + \frac{2}{(\mu\tau)^2}  - 
    \frac{ (1+i \mu \tau)^{1-N} + (1-i \mu \tau)^{1-N}  }{(\mu\tau)^2}.
\end{equation}
As we show below in Sec. S6, in the thermodynamic limit, i.e. $N\rightarrow \infty$ both of these limits reduce to the conjecture of Berry and Tabor for integrable systems~\cite{BerryTabor1977_level,BerryTabor_conjecture_Jens}, namely
\begin{equation}
    \lim_{N_R\rightarrow\infty}
    \lim_{N\rightarrow N_R}\frac{1}{N}K(\tau,N)= \lim_{N\rightarrow\infty}\lim_{N_R\rightarrow\infty}\frac{1}{N}K(\tau,N)=1+2\pi \delta(\tau).
\end{equation}

We conclude this section with an additional comment on taking the thermodynamic limit. In order to extract the Poisson behavior as $N\rightarrow \infty$, especially the sub leading $\frac{1}{(\mu \tau)^2}$ signature for $K_c(\tau,N)$ we must respect the order of limits: first we should take $N_R \rightarrow \infty$ \emph{and then} take $N \rightarrow \infty$. In fact, the procedure to extract Poisson statistics from model MBL Hamiltonians also needs similar care. Oganesyan and Huse~\cite{OganesyanHuse_2007_PhysRevB.75.155111} point out that given a disordered Hamiltonian with $L$
 sites, the number of random numbers used to describe the microscopic model are $\mathcal{O}(L)$. However, the number of eigenvalues produced are $\mathcal{O}(2^L)$. This means that there would be significant correlations between them. However, if we consider $N$ eigenvalues from deep within the spectrum, as we have in the main text, the Poisson nature is revealed if $N \ll \mathcal{N}_L$ where $\mathcal{N}_L$ is the total size of the Hilbert space. To observe Poisson statistics for arbitrary $N$, we need to take $L \rightarrow \infty$ first.

\section{S4. Disorder independence of the SFF in the MBL phase }
\begin{figure}[!h]
	\centering
	\begin{tabular}{c}
		\includegraphics[width=0.42\textwidth]{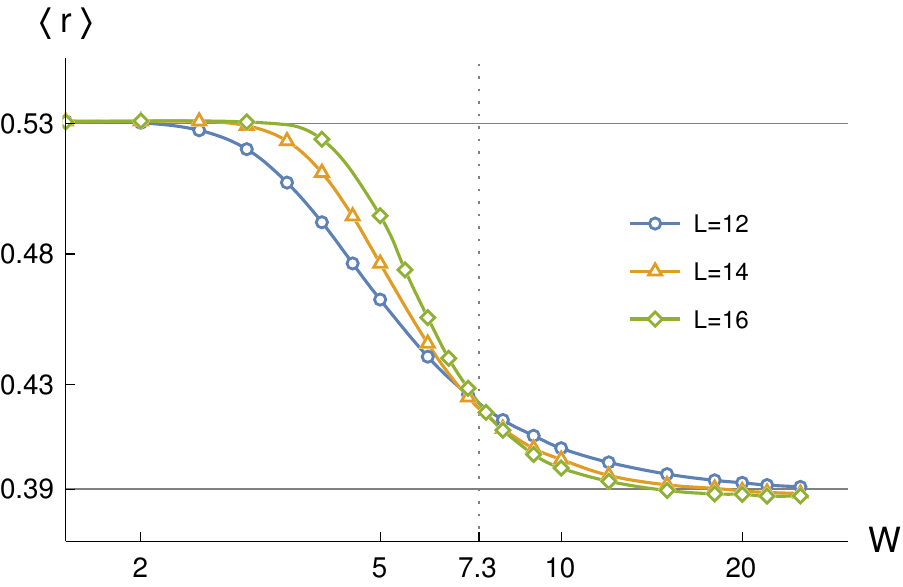} 
	\end{tabular}
	\caption{Adjacent gap ratio for the disordered Hamiltonian~\cref{eq:disorderedH} for three system sizes, showing a clear crossing in the vicinity of the MBL transition (this is shown as the inset in fig. 1 of the main text).}
	\label{fig:r_disorder}
\end{figure}

\begin{figure}[!h]
 	\centering
 	\begin{tabular}{cc}
 		\includegraphics[width=0.42\textwidth]{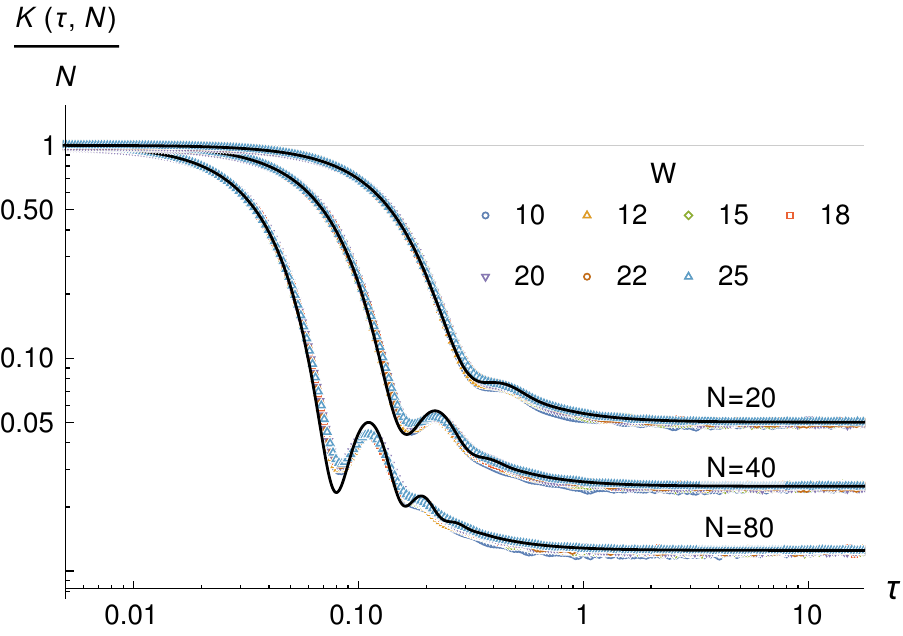} &
 		\includegraphics[width=0.42\textwidth]{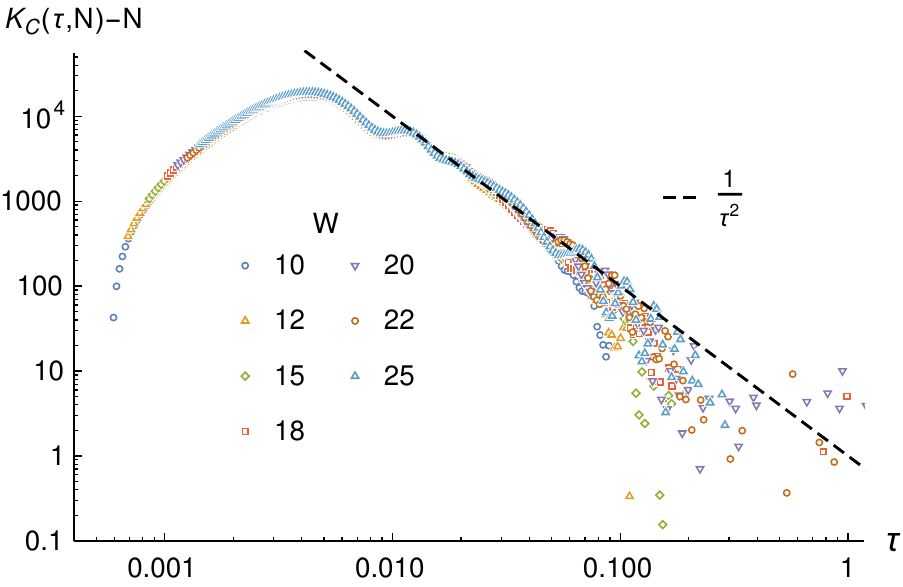}  		
 	\end{tabular}
 	\caption{For system size $L=14$, a comparison of (left) the SFF for various disorder  strengths in the MBL phase of the disordered Hamiltonian~\cref{eq:disorderedH} for small values of $N=20,40,80$ and (right) the reduced SFF for $N=900$. Up to  numerical uncertainty, we find remarkable agreement with the analytic result across all of our numerical data for $W\ge 10$.
 	}
 	\label{fig:Hd_appendix}
\end{figure}

We now focus on the disorder dependence of the SFF in the model of the disordered spin-chain that we considered in the main text. We list it here again for completeness 
\begin{align}
H(W) &= \sum_i   J_1 (S^x_i S^x_{i+1} + S^y_i S^y_{i+1} + \Delta S^z_i S^z_{i+1})+  w_i S^z_i \nonumber \\ &+  \sum_i J_2 (S^x_i S^x_{i+2} + S^y_i S^y_{i+2} + \Delta S^z_i S^z_{i+2}).
\label{eq:disorderedH}
\end{align}
As described in the main text, by keeping track of the adjacent gap  ratio $r$ that 
is defined in terms successive gaps $\delta_i = E_{i+1} - E_i$ of the energy spectrum $\{E_i\}$  as 
\begin{equation}
r_i = \frac{\mathrm{min}(\delta_i, \delta_{i+1})}{\mathrm{max}(\delta_i, \delta_{i+1})},
\label{eq:r_H}
\end{equation}
we can determine that the model above has a thermal phase for weak disorder ($W < W_c$) and an MBL phase at strong disorder ($W>W_c$). The critical disorder is roughly around $W_c \approx 7.3$ where the curves for different system sizes appear to cross as shown in \cref{fig:r_disorder}. 
In the main text, we focused on data that is deep in the MBL regime (i.e. where $\moy{r}\approx0.39$ is nicely Poisson), for a specific disorder strength $W=25$ where we showed that the SFF converges well with the analytical result. We now test the validity of this result \emph{across} the MBL phase. \Cref{fig:Hd_appendix} shows that for a wide range of disorder strengths $W \ge 10$, for small values of $N=20,40,80$ the SFF is remarkably convergent with each other and, importantly, with the analytical result. For larger values of $N=900$, the universal $1/(\mu \tau)^2$ behavior can be observed at intermediate $\tau$ values. This further strengthens the claim that MBL is indeed a robust phase.
In \cref{tab:H}, we list the number of disorder samples used for the various numerical calculations in obtaining the results presented in the main text as well as the supplementary materials. 
\begin{table}[!h]
	\centering
	\begin{tabular}{||c|| c | c | c | c ||} 
		\hline
		H(W) & L=12 & L=14 & L=16 &L=18 \\ [0.5ex] 
		\hline\hline
		W $<$ 25     & 10,000 & 10,000 & 10,000 & - \\
		W = 25     & 50,000 & 50,000 & 90,000 & 17,500 \\[1ex] 
		\hline
	\end{tabular}
	\centering
\begin{tabular}{||c|| c | c | c | c |c ||} 
	\hline
	$H_{lbit}$ & L=18 & L=20 & L=22 &L=24 & L=26 \\ [0.5ex]
	\hline\hline
	& 50,000 & 50,000 & 50,000 & 50,000 & 12,000 \\[1ex] 
	\hline
\end{tabular}    
	\caption{Number of disorder samples analyzed in studying (left) the disordered spin chain model~\cref{eq:disorderedH} and (right) the truncated $l$-bit model~\cref{eq:truncatedLbit}}
	\label{tab:H}
\end{table}

\section{S5. Quantifying the SFF deviations from exact result}
\begin{figure}[!h]
	\centering
	\begin{tabular}{ccc}
		\includegraphics[width=0.32\textwidth]{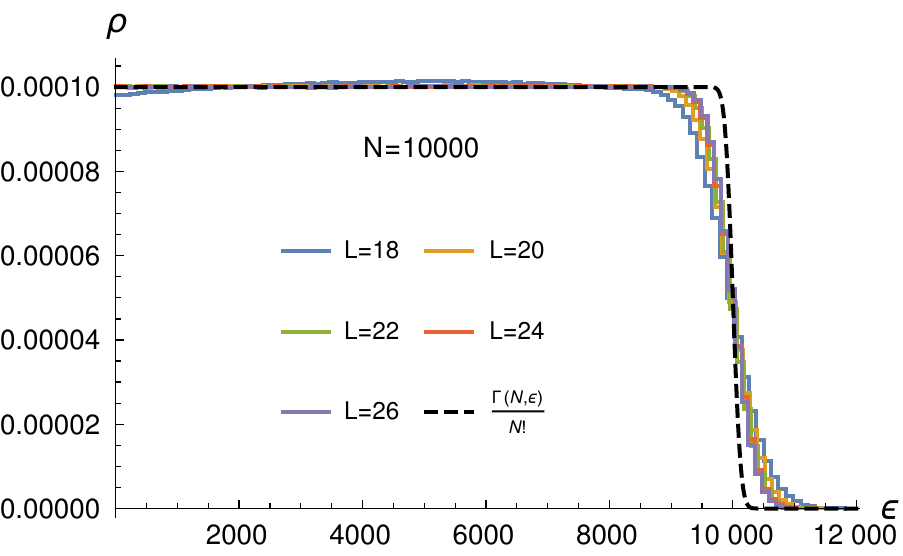} &
		\includegraphics[width=0.32\textwidth]{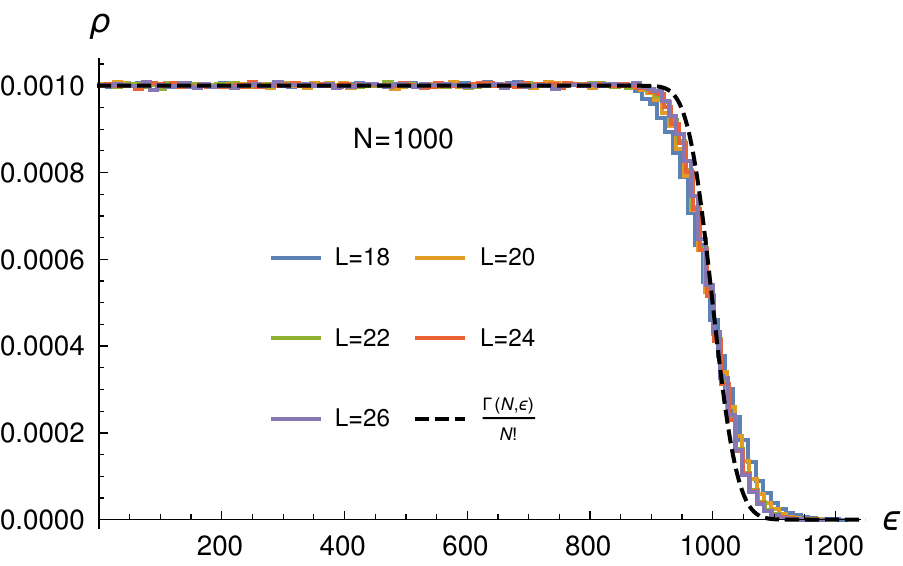} &
		\includegraphics[width=0.32\textwidth]{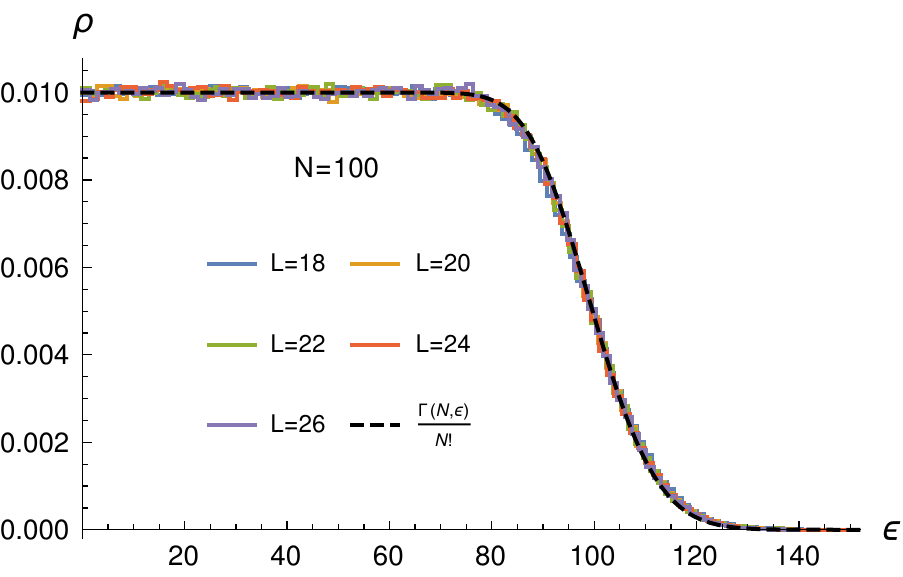}\\
		\includegraphics[width=0.32\textwidth]{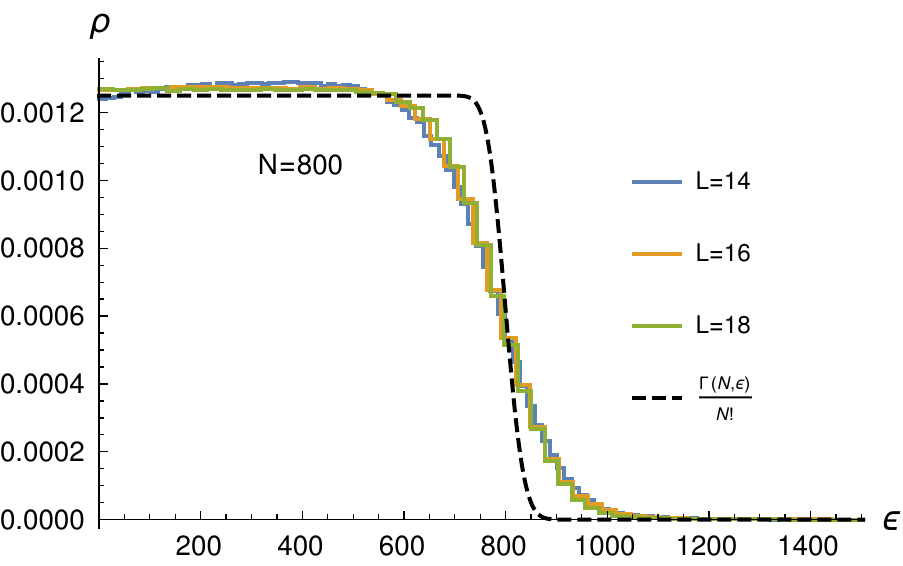} &
		\includegraphics[width=0.32\textwidth]{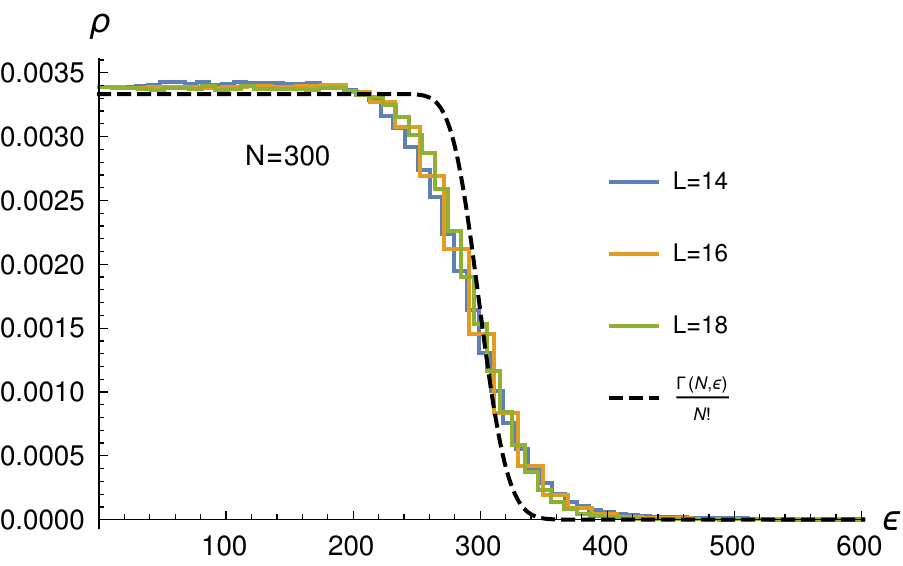} &
		\includegraphics[width=0.32\textwidth]{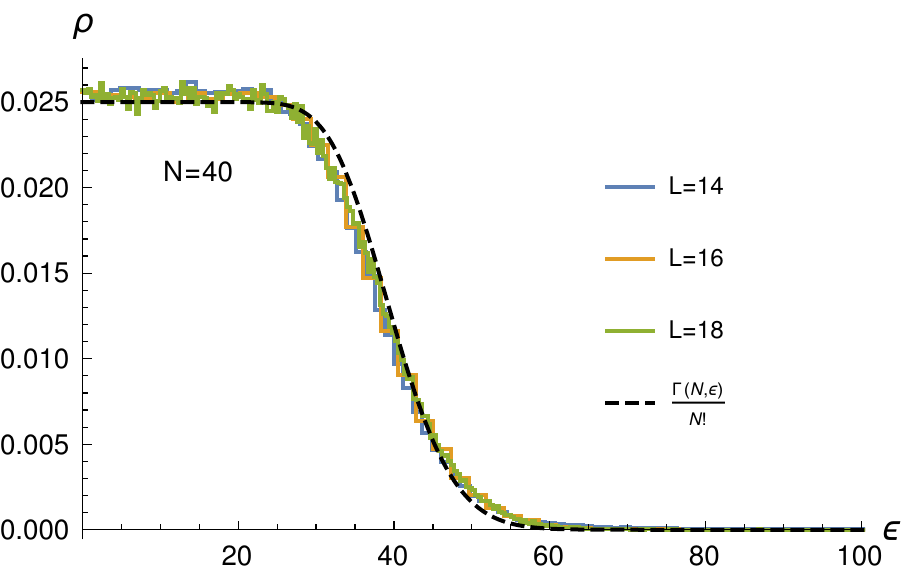}  
	\end{tabular}
	\caption{The many-body density of states $\rho(E)$ for the truncated $l$-bit model~\cref{eq:truncatedLbit} (top) with the disordered Hamiltonian~\cref{eq:disorderedH} with $W=25$ (bottom) for different choices for eigenvalues $N$ in the $S^z=0$ sector, $\mathcal{N}_L = \binom{L}{L/2}$ for various system sizes $L$, averaged over 10000 disorder samples. The black dashed line is the expectation for a Poisson density of states given in Eq.~\eqref{eq:Poisson_DOS}, which agrees well with our numerical results in the limit $N/\mathcal{N}_L\rightarrow 0$.}
	\label{fig:hist}
\end{figure}

\begin{figure}[!h]
	\centering
	\begin{tabular}{cc}
		\includegraphics[width=0.42\textwidth]{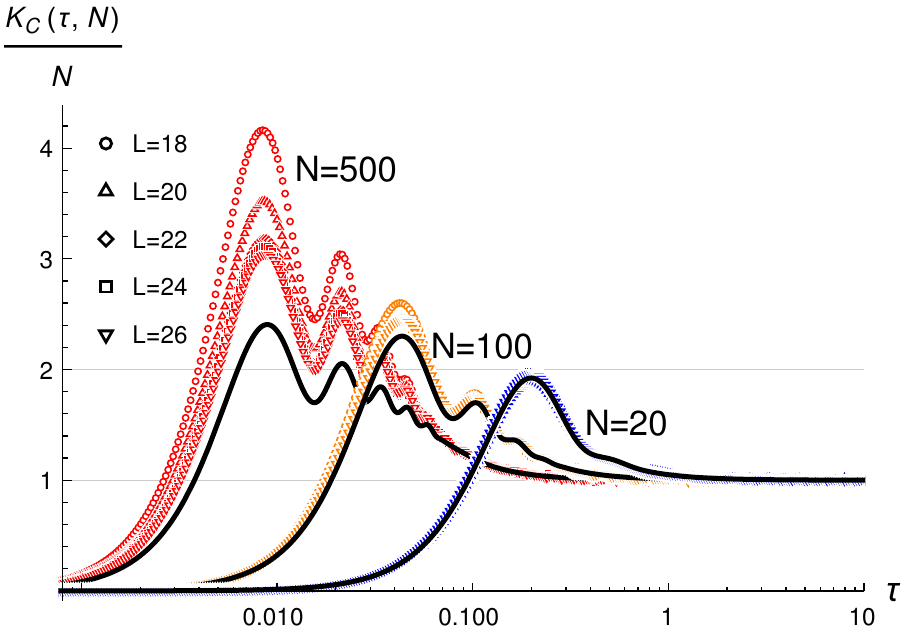} &
		\includegraphics[width=0.42\textwidth]{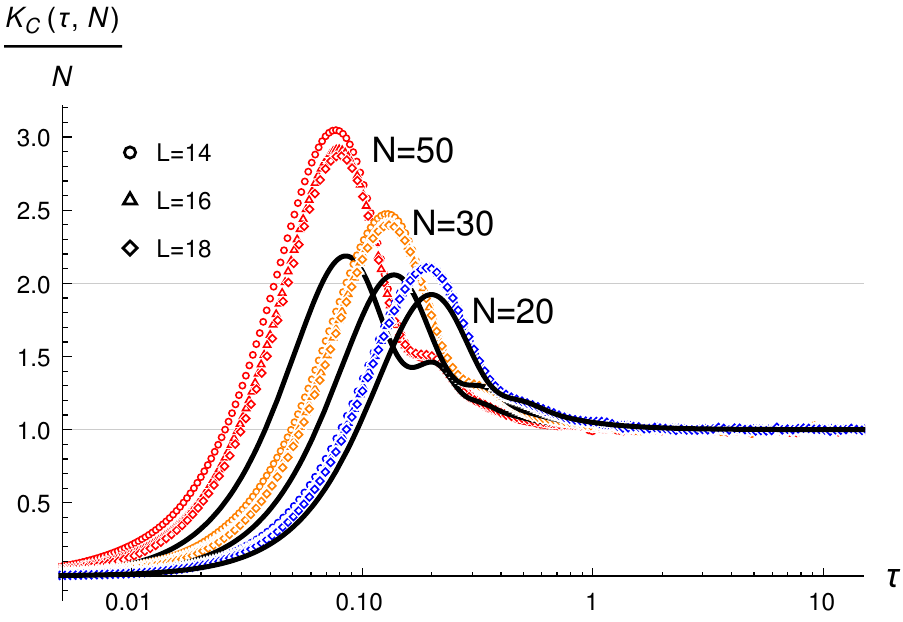} 
	\end{tabular}
	\caption{Comparison of the analytic result for the SFF of Poisson levels for various values of $L$ and $N$ with the numerics for the $l$-bit model (left) and the disordered Hamiltonian (right).  }
	\label{fig:sffc}
\end{figure}

\begin{figure}[!h]
	\centering
	\begin{tabular}{cc}
		\includegraphics[width=0.42\textwidth]{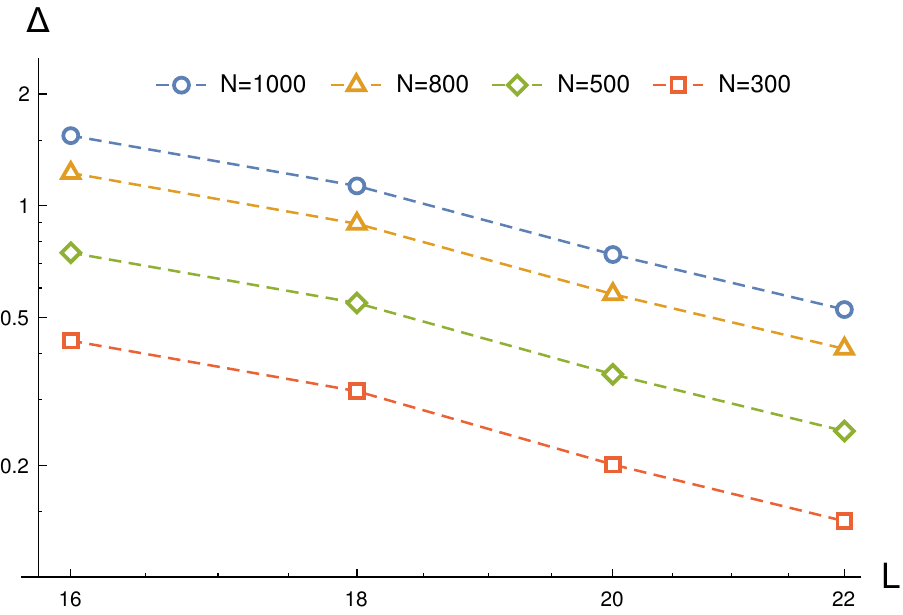} &
		\includegraphics[width=0.42\textwidth]{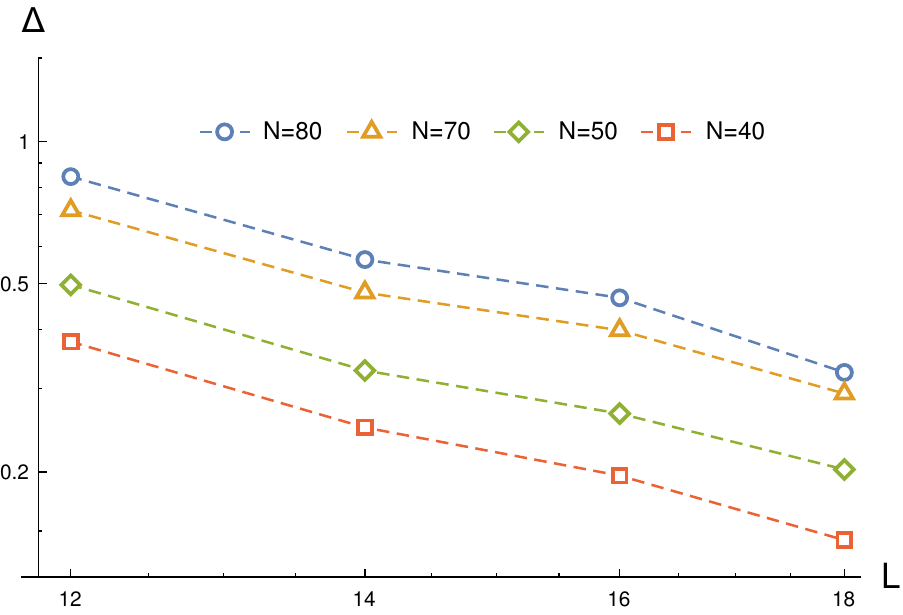} 
	\end{tabular}
	\caption{Root mean square deviation of the connected SFF from analytical result for $l$-bit (left) and Hamiltonian (right).}
	\label{fig:rms}
\end{figure}

At finite sizes, deviations from the exact result due the deviation of the many-body density of states (DOS) from the Poisson form are inevitable at early values of $\tau$. This is minimized by picking out $N$ eigenvalues deeper in the middle of the spectrum where the DOS is closest to Poisson. To examine this, we compute the many-body DOS defined as 
\begin{equation}
    \rho(E) =\frac{1}{N}\sum_{i=1}^N\delta(E - E_i),
\end{equation} 
in comparison with the DOS of Poisson numbers. The latter can be computed exactly as follows
\begin{align}
\moy{\rho(E)} &= \frac{1}{N} \sum_{k=1}^N \moy{\delta(E - E_k)} = \frac{1}{N}  \sum_{k=1}^N \int_0^\infty dE P(k,E_k)~\delta(E - E_k) \nonumber \\
&= \frac{1}{N } \sum_{k=1}^N \left(\frac{E}{\mu} \right)^{k-1} \frac{e^{-\frac{E}{\mu}}}{\mu(k-1)!} = \frac{\Gamma\left(N,E/ \mu \right)}{\mu~N!}, \label{eq:Poisson_DOS}
\end{align}
where, $\Gamma(N,x)$ is the \emph{incomplete Gamma function} defined (for integer $N$) as
\begin{align}
\Gamma(N,x) = \int_x^\infty dt~t^{N-1} e^{-t} = e^{-x}~ (N-1)!  \sum_{k=0}^N \frac{x^k}{k!}.
\end{align}
As shown in \cref{fig:hist},
we observe that for $N$ corresponding to increasingly smaller fractions of  the total Hilbert space dimension in the $S^z=0$ sector, $\mathcal{N}_L = \binom{L}{L/2}$, the DOS increasingly approaches the analytical form computed for Poisson numbers~\cref{eq:Poisson_DOS} for both the disordered spin chain model $H(W)$ of~\cref{eq:disorderedH} in the MBL phase ($W=25$) as well as the truncated $l$-bit model~\cref{eq:truncatedLbit} (we list here again the $l$-bit model for completeness). 
\begin{align}
H_{lbit} &= \sum_{i=1}^L \sum_{a=1}^{10} J^a_i \kappa^z_i \kappa^z_{i+1} \ldots \kappa^z_{i+a} \label{eq:truncatedLbit}
\end{align}
Despite minimizing this deviation by selecting the eigenvalues from the middle of the many-body spectrum, differences are seen at the edge of the selected eigenvalues. The SFF results at small $\tau$ values are dominated by large energy differences, which probe the edges of the many-body spectrum where there are bound to be deviations at finite sizes. On the other hand, for intermediate values of $\tau$, the SFF probes the eigenvalues at smaller scales where the DOS of the Poisson numbers is flat and we should see good agreement with theoretical expectations, including the $\frac{1}{\tau^2}$ behavior. 
The DOS curvature and thus the deviation from the analytical result are expected to reduce by increasing system size ($L$ and thus the size of the Hilbert space $\binom{L}{L/2}$) and reducing $N$. This is clearest in the connected SFF as shown in \cref{fig:sffc}. Consistent with the expectations stated above, we find the largest area of deviation occurs at short $\tau$ values, where as larger $\tau$ values (but well before the plateau sets in) that are sampling energy differences from the flat DOS, our analytic results match the numerics nicely.
To quantify the deviation, we consider the quantity $\Delta$ defined as the root mean square  difference of the numerical connected SFF from the analytical value at various values of $\tau$ defined as
\begin{eqnarray}
\Delta =  \sqrt{ \frac{\sum_{i=1}^{N_\tau}(K_C(\tau_i)-K^{P}_C(\tau_i))^2}{N_\tau} },
\end{eqnarray}
$\tau_i$ is chosen from the interval $\tau_i \in (10^{-5},15.0)$ (with mean-level spacing set to 1) where the SFF is non-trivial and $N_\tau$ is the total number of $\tau$ points sampled. As seen in \cref{fig:rms}, we see $\Delta$ decrease as expected with increasing $L$ as well as decreasing $N$.

\section{S6. Connection to the Berry Tabor conjecture and the $N\rightarrow \infty$ limit}

\label{sec:bt}

\subsection{S6a. Review of the statement of the conjecture}
Berry and Tabor~\cite{BerryTabor1977_level,BerryTabor_conjecture_Jens} were interested in the spectra of point particles which have classically integrable geodesics. For examples, free particles in a manifold $\cM$ with positive curvature such as spheres and tori or with boundary conditions such as circular or rectangular. The spectrum is basically that of the Schrodinger operator which is  
\begin{align}
-\frac{\hbar^2}{2m} \triangle \phi_i = \lambda_i \phi_i
\end{align}
where $\triangle$ is the Laplace-Beltrami operator on $\cM$ and the eigenvalues $\lambda_i$ are positive definite, countably infinite and unbounded from above $0 < \lambda_1 < \lambda_2 < \ldots \rightarrow \infty $. The question that the authors were interested is whether the distribution of spectral gaps assumes an asymptotic form. Specifically, if we define
\begin{align}
P(s,N) = \frac{1}{N} \sum_i \delta(s-\lambda_{i+1}+ \lambda_i),
\end{align}
the question is whether the following limit exists and what its form is
\begin{equation}
P(s,N) \xrightarrow{N\rightarrow \infty } P(s).
\end{equation}
Berry and Tabor conjecture that the limit does exist for sufficiently generic systems and corresponds to that of a Poisson process
\begin{align}
P(s) = \frac{1}{\mu} e^{-\frac{s}{\mu}}.
\end{align}
The Berry Tabor conjecture is sometimes also stated in terms of the pair density correlator
\begin{align}
R_2(s) = \frac{1}{N} \sum_{i,j=1}^N \delta(s-E_{j}+E_{i}) \xrightarrow{N \rightarrow \infty}  \delta(s) + 1,
\end{align}
and its Fourier transform, the spectral form factor
\begin{align}
\frac{K(\tau,N)}{N} = \frac{1}{N}\sum_{m,n =1}^N \langle e^{i \tau (E_m - E_n)} \rangle_P   \xrightarrow{N \rightarrow \infty}  1 + 2 \pi \delta(\tau). \label{eq:Poisson limit}
\end{align}

\subsection{S6b. Matching the $N \rightarrow \infty$ limit for the Poisson SFF}
\label{app:BT_Ninfinity}
\begin{figure}[!h]
	\centering
	\begin{tabular}{c}
		\includegraphics[width=0.5\textwidth]{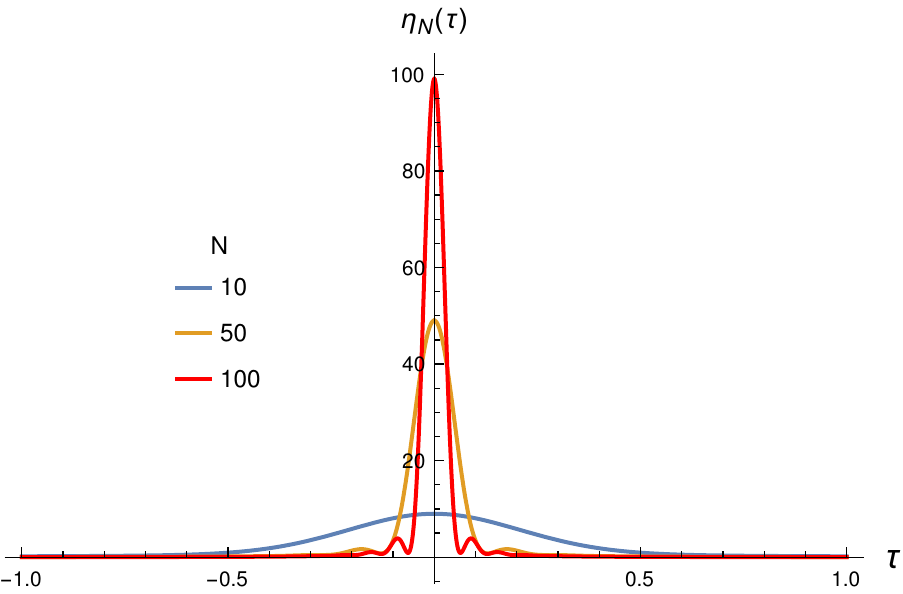} 
	\end{tabular}
	\caption{$\eta_N(\tau)$ for various $N$ \label{fig:PoissontoDelta}}
\end{figure}

A consequence of the $l-bit$ hypothesis states that the MBL spectrum, due to emergent integrability should also satisfy the Berry-Tabor conjecture. This is supported by various measures of spectral correlations and statistics being reproduced by Poisson numbers. We now show that the spectral form factor of \cref{eq:SFF_Poisson2} reduces to \cref{eq:Poisson limit} in the appropriate limit. Let us start with \cref{eq:SFF_Poisson2} (setting $\mu=1$ for convenience)
\begin{align}
K^P(\tau,N) &= N + \frac{2}{\tau^2}  - \frac{ (1+i  \tau)^{1-N} + (1-i \tau)^{1-N}  }{\tau^2}.
\end{align}
We wish to recover the limit for integrable models~\cref{eq:Poisson limit}
\begin{equation}
\lim_{N \rightarrow \infty}  \frac{K^P(\tau,N)}{N} =  1 + 2 \pi \delta(\tau).
\end{equation}
This amounts to proving that 
\begin{align}
\lim_{N \rightarrow \infty} \eta_N(\tau) &= 2 \pi \delta (\tau) \\
\text{ where, } \eta_N(\tau) &= \frac{2}{N\tau^2}  - \frac{ (1+i  \tau)^{1-N} + (1-i \tau)^{1-N}  }{N\tau^2} \label{eq:eta}
\end{align}
It is very clear from the plots of $\eta_N(\tau)$~\cref{fig:PoissontoDelta} that the functional form looks increasingly like the delta function with increasing $N$. 
To prove the identity~\cref{eq:Poisson limit}, we need to find the limiting form of $K^P(\tau,N)$ when viewed as an integration measure i.e. we want to show that, for $f(\tau)$ that is well-behaved on the real line $\tau \in \bR$, we need to show that 
\begin{align}
\lim_{N \rightarrow \infty} \int_{-\infty}^{\infty} d \tau f(\tau) \eta_N(\tau) = 2 \pi f(0). \label{eq: delta limit general}
\end{align}
To do this, we need to prove that $\eta_N(\tau)$ satisfies three conditions~\cite{balakrishnan2003all}:
\begin{enumerate}
	\item Symmetry: $ \eta_N(-\tau) = \eta_N(\tau)$
	\item Singularity: $\lim_{N \rightarrow \infty} \frac{\eta_N(\tau \neq 0)}{\lim_{\tau \rightarrow 0} \eta_N(\tau)} = 0$
	\item Normalization: $\lim_{N \rightarrow \infty} \int d\tau \eta_N(\tau) = 2 \pi$ 	
\end{enumerate}

The first i.e. symmetry is verified easily from the form of $\eta_N(\tau)$ in \cref{eq:eta}. Singularity, which ensures that the function is increasingly peaked at $\tau =0$ is also verified in a straight forward manner:
\begin{align}
\lim_{\tau \rightarrow 0} \eta_N(\tau) &= N-1 \\
\lim_{N \rightarrow \infty} \frac{\eta_N(\tau \neq 0)}{\lim_{\tau \rightarrow 0} \eta_N(\tau)} &=\lim_{N \rightarrow \infty} \frac{2}{N(N-1)\tau^2}  - \frac{ (1+i  \tau)^{1-N} + (1-i \tau)^{1-N}  }{N(N-1)\tau^2} = 0 
\end{align}
The last i.e. normalization requires some work. We need to compute the area under the curve $\eta_N(\tau)$. We evaluate the integral using tricks used to evaluate \emph{Dirichlet integrals} starting by splitting it up as follows
\begin{align}
\int_{-\infty}^{\infty} d\tau \eta_N(\tau) &= \cP \left( \int_{-\infty}^{\infty} d\tau I_+(N,\tau) \right) + \cP \left( \int_{-\infty}^{\infty} d\tau I_-(N,\tau) \right), \\
\text{ where, } I_\pm(N,\tau) &= \frac{1- (1+i \tau)^{1-N}}{N \tau^2}.
\end{align}
Note that while $\eta_N(\tau)$ is well behaved on the real line, $I_{\pm}$ are individually singular at $\tau = 0$. $\cP$ refers to the principal component integration defined as follows
\begin{equation}
\cP \left( \int_{-\infty}^{\infty} d\tau I_\pm(N,\tau) \right) = \lim_{\epsilon \rightarrow 0} \left(\int_{-\infty}^{-\epsilon} d\tau I_\pm(N,\tau) + \int_{\epsilon}^{\infty} d\tau I_\pm(N,\tau)\right).
\end{equation}
Let us proceed to evaluate each piece of the integral starting with $I_+$. Notice that $I_+$ is singular at two points on the complex plane- $\tau=0$ and $\tau = i$.  
\begin{figure}[!h]
	\centering
	\begin{tabular}{c}
		\includegraphics[width=0.5\textwidth]{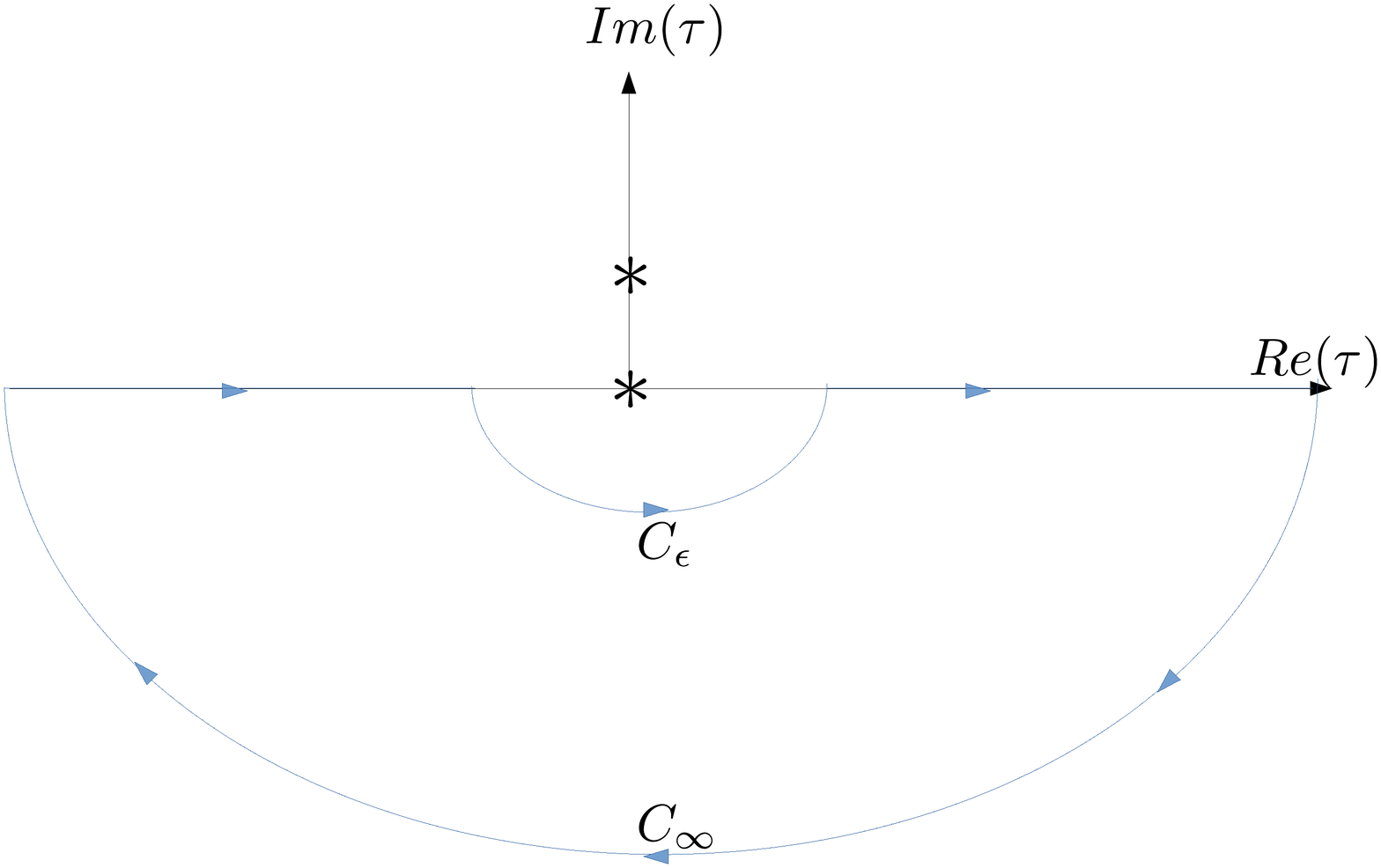} 
	\end{tabular}
	\caption{Contour $C$ to evaluate the $I_+$ integral. The poles of $I_+$ are marked as $*$ \label{fig:contour 1}}
\end{figure}
Consider a contour that avoids both these points as shown in \cref{fig:contour 1}. $I_+(N,\tau)$ is analytic at all points inside the contour and hence, by Cauchy's theorem, we get
\begin{align}
\varointclockwise_C dz I_+(N,z) = 0.
\end{align}
We can split the integral on the above contour as
\begin{align}
\int_{-\infty}^{-\epsilon} d\tau I_+(N,\tau)+ \int_{\epsilon}^\infty d\tau I_+(N,\tau) + \int_{C_\epsilon} dz I_+(N,z) + \int_{C_\infty} dz I_+(N,z) =0 
\end{align}
The contribution of the integral along the arc at infinity is $0$ since the integrand $I_+$ vanishes at infinity. If we take the limit of the radius of the semicircle encircling the origin, $\epsilon \rightarrow 0$, we have the following
\begin{align}
\cP \left( \int_{-\infty}^{\infty} d\tau I_+(N,\tau) \right) = - \lim_{\epsilon \rightarrow 0} \int_{C_\epsilon} dz I_+(N,z). 
\end{align}
The RHS of the above equation is evaluated easily. Along the semi-circular arc $C_\epsilon$, the complex numbers can be parametrized as $z = \epsilon~ e^{i \theta}$. Thus we get
\begin{align}
\cP \left( \int_{-\infty}^{\infty} d\tau I_+(N,\tau) \right) &= - \lim_{\epsilon \rightarrow 0} \int_{-\pi}^0 i d \theta ~(\epsilon~ e^{i \theta}) \frac{1- (1+ i \epsilon~ e^{i \theta})^{1-N}}{N (\epsilon~ e^{i \theta})^2} \nonumber \\&\approx - \lim_{\epsilon \rightarrow 0} \int_{-\pi}^0 i d \theta ~(\epsilon~ e^{i \theta}) \frac{1- (1+ i (1-N) \epsilon~ e^{i \theta})}{N (\epsilon~ e^{i \theta})^2} \nonumber \nonumber = \pi \left(1- \frac{1}{N}\right).
\end{align}
The above steps can be repeated for $I_-$ which has poles at $0,-i$ and can be evaluated by closing the contour on the upper half complex plane to get
\begin{align}
\cP \left( \int_{-\infty}^{\infty} d\tau I_-(N,\tau) \right) &= \pi \left(1- \frac{1}{N}\right).
\end{align}
Putting all these together, we get 
\begin{align}
\int_{-\infty}^\infty d \tau \eta_N(\tau) &= 2 \pi \left(1- \frac{1}{N}\right), \\
\lim_{N \rightarrow \infty} \int_{-\infty}^\infty d \tau \eta_N(\tau) &= 2 \pi .
\end{align}
Using these, for any well-behaved test function $f(\tau)$, we can conclude
\begin{align}
\int_{-\infty}^\infty d \tau f(\tau)  \eta_N(\tau) \approx  f(0) \int_{-\infty}^\infty d \tau \eta_N(\tau) + \mathcal{O} \left(\frac{1}{N}\right)  = 2 \pi f(0) + \mathcal{O} \left(\frac{1}{N}\right).
\end{align}
Taking the limit of $N \rightarrow \infty$ above, we get \cref{eq: delta limit general}.

\subsection{S6c. Matching the $N \rightarrow \infty$ limit for the Random SFF}
\begin{figure}[!h]
	\centering
	\begin{tabular}{c}
		\includegraphics[width=0.5\textwidth]{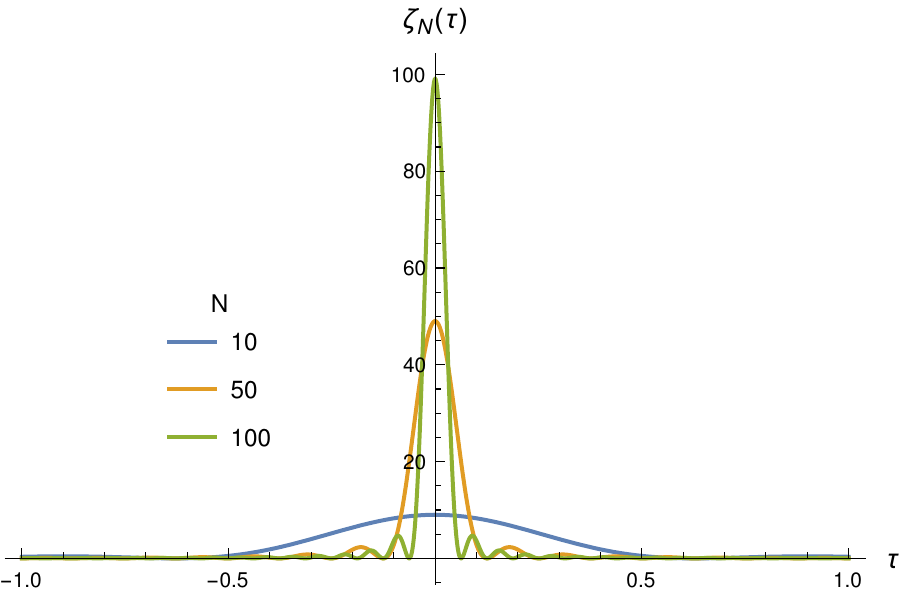} 
	\end{tabular}
	\caption{$\zeta_N(\tau)$ for various $N$ \label{fig:zeta}}
\end{figure}

We now obtain the same $N \rightarrow \infty$ limit for the SFF of uniformly distributed random numbers and show that \cref{eq:Uniform SFF} reduces to \cref{eq:Poisson limit}. Let us start with eq~\ref{eq:Uniform SFF} setting $\mu=1$ for convenience as before.
\begin{align}
K^R(\tau,N) &= N + N(N-1) \left| \frac{\sin(N \tau/2)}{(N \tau/2)} \right|^2.
\end{align}
We wish to recover the limit for integrable models~\cref{eq:Poisson limit}
\begin{equation}
\lim_{N \rightarrow \infty}  \frac{K^R(\tau,N)}{N} =  1 + 2 \pi \delta(\tau).
\end{equation}
This amounts to proving that 
\begin{align}
\lim_{N \rightarrow \infty} \zeta_N(\tau) &= 2 \pi \delta (\tau) \\
\text{ where, } \zeta_N(\tau) &= (N-1) \left| \frac{\sin(N \tau/2)}{(N \tau/2)} \right|^2 \label{eq:zeta}.
\end{align}
It is very clear from the plots of $\zeta_N(\tau)$~\cref{fig:zeta} that the functional form looks increasingly like the delta function with increasing $N$.
Again, we need to show that the functional form of $\zeta_N$  approaches that of the delta function when interpreted as an integration measure i.e. it satisfies~\cite{balakrishnan2003all}:
\begin{enumerate}
	\item Symmetry: $ \zeta_N(-\tau) = \zeta_N(\tau)$
	\item Singularity: $\lim_{N \rightarrow \infty} \frac{\zeta_N(\tau \neq 0)}{\lim_{\tau \rightarrow 0} \zeta_N(\tau)} = 0$
	\item Normalization: $\lim_{N \rightarrow \infty} \int d\tau \zeta_N(\tau) = 2 \pi$ 	
\end{enumerate}
The first two are again easily verified. The third too is verified as follows. First, let us change variables $\frac{\mu \N \tau}{2} = x$ to get
\begin{align}
\int_{- \infty}^\infty d\tau \zeta_N(\tau) &= (N-1) \int_{- \infty}^\infty d\tau \left( \frac{\sin(N \tau/2)}{(N \tau/2)}  \right)^2 = \frac{2(N-1)}{N} \int_{- \infty}^\infty d\tau \left( \frac{\sin x}{x}  \right)^2 = \frac{2 \pi (N-1)}{N} \\
\implies \lim_{N \rightarrow \infty} \int d\tau \zeta_N(\tau) &= 2 \pi.
\end{align}
The integral $\int_{- \infty}^\infty d\tau \left( \frac{\sin x}{x}  \right)^2 = \pi$ can be evaluated in the following way. Let us define
\begin{align}
I(b) &= \int_{- \infty}^\infty d\tau \left( \frac{\sin (bx)}{x}  \right)^2 \label{eq:I(b)}\\
\frac{d I(b)}{d b } &=  \int_{- \infty}^\infty d\tau \left( \frac{\sin (2bx)}{x}  \right) = \pi.
\end{align}
The result of $\frac{d I(b)}{d b }$ is a standard integral that can be obtained using the Dirichlet integral trick in the previous section.  We now perform the definite integral to get
\begin{equation}
I(b) = \pi b + const.
\end{equation}
Finally, the constant is evaluated to $0$ by observing $I(0) =0$ in \cref{eq:I(b)}. Finally, setting $b=1$, we get 
\begin{equation}
I(1) = \int_{- \infty}^\infty d\tau \left( \frac{\sin x}{x}  \right)^2 = \pi.
\end{equation}
This completes the proof.

\bibliography{references}{}